\documentclass[a4paper,11pt]{article}
\pdfoutput=1 

\usepackage{jcappub} 

\usepackage[T1]{fontenc} 

\title{\boldmath Finding Horndeski theories with Einstein gravity limits}

\author[a]{Ryan McManus,}
\author[a]{Lucas Lombriser,}
\author[a]{Jorge Pe\~{n}arrubia}

\affiliation[a]{Institute for Astronomy, University of Edinburgh,\\Royal Observatory, Blackford Hill, Edinburgh, EH9 3HJ, U.K.}

\emailAdd{ryanm@roe.ac.uk}
\emailAdd{llo@roe.ac.uk}
\emailAdd{jorpega@roe.ac.uk}

\abstract{
The Horndeski action is the most general scalar-tensor theory with at most second-order derivatives in the equations of motion, thus evading Ostrogradsky instabilities and making it of interest when modifying gravity at large scales. 
To pass local tests of gravity, these modifications predominantly rely on nonlinear screening mechanisms that recover Einstein's Theory of General Relativity in regions of high density.
We derive a set of conditions on the four free functions of the Horndeski action that examine whether a specific model embedded in the action possesses an Einstein gravity limit or not.
For this purpose, we develop a new and surprisingly simple scaling method that identifies dominant terms in the equations of motion by considering formal limits of the couplings that enter through the new terms in the modified action.
This enables us to find regimes where nonlinear terms dominate and Einstein's field equations are recovered to leading order.
Together with an efficient approximation of the scalar field profile, one can then further evaluate whether these limits can be attributed to a genuine screening effect.
For illustration, we apply the analysis to both a cubic galileon and a chameleon model as well as to Brans-Dicke theory.
Finally, we emphasise that the scaling method also provides a natural approach for performing post-Newtonian expansions in screened regimes.
}

\begin{document}
\maketitle
\flushbottom

\section{Introduction}
\label{sec:intro}

The late-time accelerated expansion of our Universe has been confirmed by a wealth of observational evidence from the measured distances of type Ia supernovae~\citep{ObservationalEvidenceFromSupernovaeForAnAcceleratingUniverseAndACosmologicalConstant, MeasurementsOfOmegaAndLambdaFrom42HighRedshiftSupernovae} to measurements of the secondary anisotropies of the cosmic microwave background~\citep{Planck2015ResultsXIIICosmologicalParameters}.
The simplest explanation for the effect is the contribution of a positive cosmological constant $\Lambda$ to the Einstein field equations.
The cosmological constant is a crucial constituent of the standard model of cosmology, $\Lambda$ Cold Dark Matter ($\Lambda$CDM), where it dominates the present energy budget of the Universe.
One would expect the Planck mass $M_{p}$ to set a natural scale for $\Lambda$ as it defines the relevant scale for matter interactions in the field equations, but the measured cosmological constant, when expressed in terms of $M_{p}$, is inexplicably small, $\Lambda_{\rm obs} \approx (10^{-30} M_{p})^4$~\citep{TheQuantumVacuumAndTheCosmologicalConstantProblem}.
This calls into question whether Einstein's Theory of General Relativity (GR) with the observed small value of $\Lambda_{\rm obs}$ is a fundamental description of gravity.
One may hope for a theoretical justification for $\Lambda_{\rm obs}$ from quantum-field-theory, but calculations from Standard-Model vacuum diagrams put the expected value in the order of $\Lambda_{\rm theory} \approx (10^{-15}M_{\rm pl})^4$ by assuming an ultra-violet cut-off at the scales probed with the Large Hadron Collider~\citep{TheCosmologicalConstantProblem}.
The difference of 60 orders of magnitude begs for an explanation.

To overcome such issues, much work has been performed to expand the $\Lambda$CDM model through modifying gravity directly instead of trying to solve the problem within the standard particle and cosmological paradigms. This has often been done through the introduction of new degrees of freedom. 
Postulating the presence of an extra degree of freedom is not a new concept in cosmology and is, for instance, done to facilitate inflation in the early universe.
The most common addition is the introduction of a minimally coupled scalar field with an appropriate potential.
The discovery of the Standard-Model Higgs particle has seemingly confirmed the existence of  fundamental scalar fields \cite{ObservationOfANewBosonAtAMassOf125GeVWithTheCMSexperimentAtTheLHC}.
Furthermore, quantum corrections to the Einstein-Hilbert action give rise to an effective field theory containing terms such as $R^2$, where $R$ is the Ricci scalar; this can be shown to be equivalent to adding a non-minimally coupled scalar field~\citep{TheContentOffRGravity}.
In general, one can expect the appearance of new degrees of freedom from such effective field theory considerations because of Lovelock's theorem.
These corrections often cause non-minimal couplings of gravity to these new degrees of freedom, giving rise to a wide variety of gravity models.
As such, it seems that gravity must have a more accurate description than Einstein's theory at high energies but still well below the Planck scale, so that a quantum theory of gravity is not needed and we can use classical fields.  

The easiest modification is to consider the addition of a non-minimally coupled scalar field in the Einstein-Hilbert action,  adding only one more degree of freedom.
In general, one may also want to consider higher-derivative actions of the scalar field beyond first derivatives such as in kinetic terms.
The instinctive problem with a higher-derivative theory is that they can contain Ostrogradsky instabilities, ghost degrees of freedom due to third or higher time derivatives in the equations of motion.
To avoid it, one may require the equations of motion to be of at most second order in derivatives.
This consideration leads to the Horndeski action~\citep{SecondOrderScalarTensorFieldEquationsInAFourDimensionalSpace}, describing the most general four-dimensional, local, second-derivative theory of gravity with the only gravitational degrees of freedom being the metric and the scalar field~\citep{FromKEssenceToGeneralizedGalileons, GeneralizedGInflation}.
Higher-derivative theories arise, for instance, as effective scalar-tensor description in the decoupling limit of braneworld scenarios~\citep{4DGravityOnABraneIn5DMinkowskiSpace} or massive gravity~\citep{ResummationofMassiveGravity} (also see Refs.~\citep{BeyondTheCosmologicalStandardModel, GalileonAsALocalModificationOfGravity}).
Healthy theories beyond the Horndeski action may also be formulated~\citep{HealthyTheoriesBeyondHorndeski, TransformingGravityTheoriesBeyondTheHorndeskiLagrangian}, where the equations of motion contain third-order time derivatives but the existence of hidden constraints prevents the appearance of ghost degrees of freedom~\citep{DegenerateHigherDerivativeTheoriesBeyondHorndeski}.

Despite the theoretical justifications to expect a modification of GR in the high-energy, strong-gravity regime, so far no observations are inconsistent with the theory. Moreover, precision measurements in the Solar System put very tight constraints on potential remnant infrared (IR) deviations~\citep{TheConfrontationBetweenGeneralRelativityAndExperiment}.
A good example of this is the measured value for $\gamma_{\rm PPN}$, the ratio of the time--time and space--space components of the metric in a low-energy static expansion, with a constraint of $\left|\gamma_{\rm PPN} - 1\right| \lesssim 10^{-5}$ set by the Cassini mission~\cite{TheConfrontationBetweenGeneralRelativityAndExperiment}. 
From this, one may infer that any modification to gravity in the IR is too small to account for deviations at large scales causing effects like late-time acceleration.

However, there exist screening mechanisms such that modifications to gravity are naturally suppressed in regions of high ambient mass density such as the Solar System or galaxies on larger scales, separating the IR regimes (see Ref.~\citep{BeyondTheCosmologicalStandardModel} for a review).
As such, the local experiments of gravity set the strength of screening required but do not immediately place tight constraints on gravity on large scales.
This motivates the use of modified gravity theories in cosmology (for reviews see Refs.~\citep{ModifiedGravityAndCosmology, Cosmologicaltestsofmodifiedgravity, DarkEnergyVsModifiedGravity}).
Scalar-tensor modifications have long been considered as a potential alternative explanation to the problem of cosmic acceleration.
However, it was recently shown in Refs.~\citep{BreakingADarkDegeneracyWithGravitationalWaves, ChallengestoSelfAccelerationinModifiedGravity} that Horndeski theories cannot provide a self-acceleration genuinely different from the contribution of dark energy or a cosmological constant if gravitational waves propagate at the speed of light.
Nevertheless, the dark energy field may couple non-minimally to the metric and modify gravity.

The screening mechanisms that can suppress the effect of this coupling in high-density regions fall into two main categories: (i)~Screening by the scalar field's local value such as in chameleon~\citep{ChameleonFieldsAwaitingSurprisesForTestsOfGravityInSpace} or symmetron~\citep{ScreeningLongRangeForcesthroughLocalSymmetryRestoration} models. 
These models have a canonical kinetic term in the Einstein frame and an effective potential of the additional scalar field that depends on environment, making the field static in deep gravitational potential wells. (ii)~Derivative screening such as in the Vainshtein mechanism~\citep{ToTheProblemOfNonvanishingGravitationMass} or in k-mouflage models~\citep{KMOUFLAGEGravity}, where derivatives of the scalar field dominate the equations of motion.
Both classes of screening mechanisms rely on nonlinear terms in the equations of motion.
When these are dominant, they cause the effect of the scalar field on the metric to become sub-dominant and hence suppress the impact on the motion of matter (when matter is minimally coupled to the metric, which we shall assume throughout the paper).
Gravitational models that employ these screening mechanisms can reduce to GR in the Solar System, passing the stringent local constraints, while yielding significant modifications of gravity on large, cosmological scales.
A further, linear shielding mechanism can additionally cause these large-scale modifications to cancel~\citep{ClassifyingLinearlyShieldedModifiedGravityModelsInEffectiveFieldTheory, BreakingADarkDegeneracyWithGravitationalWaves}.

Importantly, in screened regions, gravitational modifications from scalar field contributions in the action can still cause higher-order corrections to GR that can be tested against observations.
However, a linear expansion of the scalar field cannot describe these corrections as the nonlinear terms that give rise to the screening become should remain small in the series.
Hence, higher-order deviations from the expansion of GR in the low-energy static limit should not be used to describe the expansion of a screened theory.
To correctly describe the screened regime, a perturbative expansion can be conducted using a dual Lagrangian instead, which can be obtained from the original Lagrangian with a Legendre transformation~\citep{ClassicalDualsOfDerivativelySelfCoupledTheories} or using Lagrange multipliers~\citep{ClassicalDualsLegendreTransformsAndTheVainshteinMechanism}.
While the dual Lagrangian describes the same physics, its equations of motion allow for a perturbative series that is valid in the nonlinear regime of the original Lagrangian.
These procedures are mathematically involved and may not be suited for an application to the large variety of gravity theories or a generic gravitational action.

In this paper, we present an alternative, scaling method that finds a perturbative expansion for gravity theories in nonlinear regimes but does not rely upon finding a dual Lagrangian.
We demonstrate the operability of this method on a galileon and chameleon model, whereby we easily recover some known results for the Vainshtein and chameleon mechanisms.
We then apply the approach to the full Horndeski action and derive a number of conditions on the modifications that guarantee the existence of a limit where Einstein's field equations are recovered.
We use these results to examine several known gravity theories and their different limits.
Finally, we complement the scaling method with a technique that enables an efficient approximation of the scalar field's radial profile for a symmetric mass distribution,
which is used to assess whether the Einstein gravity limit obtained reflects a screening mechanism, where the recovery of GR holds near a massive body but not far from it.

The outline of the paper is as follows.
In section~\ref{sec:HornGravity}, we revisit the Horndeski action and cast the equations of motion in a form that is more useful to the application of the scaling method.
We introduce the novel method in section~\ref{sec:MyMethod}, where we also provide examples using both a galileon and a chameleon model to demonstrate its applicability.
In section \ref{sec:GRHorndeski}, we then use the method to derive the conditions on the free Horndeski functions that ensure the existence of an Einstein gravity limit.
We apply our findings to a range of different models to illustrate their screening effects
with an approximation of the radial scalar field profile.
We close with a discussion of our results and an outlook of their application to observational tests of gravity in section~\ref{sec:conclusions}.
For completeness, we provide the Horndeski field equations in the appendix, where we also present an alternative, coordinate-dependent scaling method.

\section{Horndeski gravity}
\label{sec:HornGravity}

The effective four-dimensional scalar-tensor theory of a string-theory-inspired braneworld scenario~\citep{4DGravityOnABraneIn5DMinkowskiSpace} that self-accelerates~\citep{CosmologyOnABraneInMinkowskiBulk} was observed to contain a second derivative of the scalar field in the action. 
Na\"ively this would yield problematic third time derivatives in the equations of motion, but instead it was found to only yield second time derivatives, therefore avoiding ghosts due to any Ostrogradsky instabilities. 
It was later noted, however, that the self-accelerted branch of the model is perturbatively not stable~(e.g., \citep{Ghostsintheselfacceleratingbraneuniverse, MoreOnGhostsInTheDvaliGabadazePorratiModel}).
Furthermore, the extra gravitational force exerted from the scalar field near massive bodies was shown to be suppressed with respect to the Newtonian force~\citep{ClassicalAndQuantumConsistencyOfTheDGPModel}.
Another interesting aspect of the effective action is its Galilean symmetry: invariance under the transformation of the field $\phi(x) \to \phi(x) + a + b_\mu x^\mu$ for constants $a$ and $b_\mu$. 
This motivated the extension of the action to the most general scalar-tensor theories invariant under this symmetry in flat space, dubbed galileon gravity~\citep{GalileonAsALocalModificationOfGravity}.
Note that this symmetry is of relevance to the non-renormalisation theorem~\citep{StrongInteractionsAndStabilityInTheDGPModel}.

In four dimensions there are five flat-space galileon Lagrangians with three of them containing higher derivatives of the scalar field but still yielding second-order equations of motion invariant under the symmetry.
Should one try to na\"ively covariantise the flat-space Lagrangians by making the metric dynamical and promoting partial to covariant derivatives, the resulting equations of motion would produce third-order time derivatives.
Ref.~\citep{CovariantGalileon} showed that counterterms consisting of non-minimal couplings between the scalar field and the metric remove the higher derivatives in the equations of motion, but at the expense of explicitly breaking the Galilean symmetry in these actions due to the couplings to gravity.
It has been shown that the breaking caused by interactions with gravity can be considered weak~\citep{WeaklyBrokenGalileonSymmetry}, so that aspects of the quantum properties of flat space galileons can be retained.
A construction of the most general scalar-tensor theory with a single scalar field on a curved space-time, not adhering to the Galilean shift symmetry, and with the derivatives in the equations of motion being of second order at most was performed in Ref.~\citep{FromKEssenceToGeneralizedGalileons}.
The resulting action was found to be equivalent~\citep{GeneralizedGInflation} to the scalar-tensor action derived earlier by Horndeski~\citep{SecondOrderScalarTensorFieldEquationsInAFourDimensionalSpace}.
Meanwhile, it was shown that the covariantisation of the flat-space galileons, while introducing third derivatives, does not necessarily yield any Ostrogradsky instabilities, which allows to formulate healthy theories beyond Horndeski gravity~\citep{HealthyTheoriesBeyondHorndeski} through evading the non-degeneracy conditions of the Ostrogradsky theorem \cite{DegenerateHigherDerivativeTheoriesBeyondHorndeski}.

We briefly review the equations of motion produced by the Horndeski action, whereby we focus on a strongly condensed form following the notation of Ref.~\citep{GeneralizedGInflation}.
The full expressions can be found in appendix~\ref{app:HornFieldEqs}.
We then point out an important pattern that appears in these equations that will become very useful in section~\ref{sec:GRHorndeski} to reduce the number of terms that need to be considered when exploring limits wherein a theory recovers Einstein gravity.

The Horndeski action, including a minimally coupled matter action $S_m[g]$, is given by
\begin{align}
\label{eq:HornAction}
S_\emph{H} =  & \frac{M_p^2}{2} \int d^4x \sqrt{-g} \bigg\lbrace G_2(\phi,X) - G_3(\phi,X) \Box \phi  \nonumber \\
&+G_4(\phi,X) R + G_{4X}(\phi,X) [ (\Box \phi)^2 - (\nabla_\mu \nabla_\nu\phi)^2 ] \nonumber \\
&+G_5(\phi,X) G_{\mu \nu}\nabla^\mu \nabla^\nu \phi - \frac{G_{5X}(\phi,X)}{6}[(\Box \phi)^3 - 3\Box \phi(\nabla_\mu \nabla_\nu \phi)^2 + 2(\nabla_\mu \nabla_\nu \phi)^3] \bigg\rbrace \nonumber \\
&+ S_m[g] \,,
\end{align}
where $G_{\mu\nu} \equiv R_{\mu\nu} - \frac{1}{2} g_{\mu\nu} R$ is the Einstein tensor, $G_2$, $G_3$, $G_4$ and $G_5$ are free functions of a scalar field $\phi$ and $X \equiv -\frac{1}{2} \partial^\mu \phi \partial_\mu \phi$, and subscripts of $X$ or $\phi$ denote functional derivatives with respect to $X$ or $\phi$.
Variation of the action with respect to the metric and scalar field yields the equations of motion~\citep{GeneralizedGInflation}
\begin{align}
\sum_{i=2}^5 \mathcal{G}_{\mu\nu}^{(i)} & = \frac{T_{\mu\nu}}{M_p^2} \,, \label{eq:PreHorndeskiMetricEom} \\
\sum_{i=2}^5 \nabla^\mu J^{(i)}_\mu & = \sum_{i=2}^5 P_\phi^{(i)} \,, \label{eq:PreHorndeskiScalarEom}
\end{align}
respectively, where $\mathcal{G}^{(i)}$, $J^{(i)}_\mu$, $P_\phi^{(i)}$ are functions of $G_i$, their derivatives and functional derivatives with respect to $\phi$ and $X$ (see Ref.~\citep{GeneralizedGInflation} or appendix \ref{app:HornFieldEqs} for $J_\mu^{(i)}$ and $P_\phi^{(i)}$), and $T_{\mu\nu}$ is the stress-energy tensor from the matter action.

Note that $G_4 G_{\mu\nu}$ appears within $\mathcal{G}^{(4)}_{\mu\nu}$ and $-G_{5\phi}XG_{\mu\nu}$ appears in $\mathcal{G}^{(5)}_{\mu\nu}$. We define $\overline{\mathcal{G}}^{(4)}_{\mu\nu} \equiv \mathcal{G}^{(4)}_{\mu\nu} - G_4 G_{\mu\nu}$ and $\overline{\mathcal{G}}^{(5)}_{\mu\nu} \equiv \mathcal{G}^{(5)}_{\mu\nu} + G_{5\phi}XG_{\mu\nu}$ as well as a $\Gamma \equiv G_4 - G_{5\phi}X$. One can then find the trace-reversed form of eq.~\eqref{eq:PreHorndeskiMetricEom} to get the metric field equations
\begin{equation}
\label{eq:HorndeskiMetricEq}
\Gamma R_{\mu\nu} = - \sum_{i=2}^5 R^{(i)}_{\mu\nu} + (T_{\mu\nu} - \frac{1}{2} g_{\mu\nu} T)/M_p^2 \,,
\end{equation}
where for convenience, we have also defined the trace-reversed tensors
\begin{align*}
R^{(i)}_{\mu\nu} &\equiv \mathcal{G}^{(i)}_{\mu\nu} - \frac{1}{2} g_{\mu\nu} g^{\alpha\beta}\mathcal{G}^{(i)}_{\alpha\beta} \,, \\
R^{(j)}_{\mu\nu} &\equiv \overline{\mathcal{G}}^{(j)}_{\mu\nu} - \frac{1}{2} g_{\mu\nu} g^{\alpha\beta}\overline{\mathcal{G}}^{(j)}_{\alpha\beta} \,, 
\end{align*}
for $i=2,3$ and $j=4,5$. Note that for $R^{(i)}_{\mu\nu}=0$ $\forall i$ and a constant $\Gamma$, the metric field equations~\eqref{eq:HorndeskiMetricEq} reduce to Einstein's equations.

We wish to source the scalar field equation~\eqref{eq:PreHorndeskiScalarEom} by the matter density.
To do this, we select the terms: 
$RG_{4\phi}$ in $P^{(4)}_\phi $, 
$-RG_{4X} \Box \phi$ in $\nabla^\mu J_\mu^{(4)} $, 
$\frac{1}{2}R G_{5X} (\Box \phi)^2 $ and $G_{5\phi} R \Box \phi$ both in $\nabla^\mu J_\mu^{(5)}$. 
These identify all possible contributions where the Ricci scalar enters the scalar field equation. 
We define $\overline{P^{(i)}_\phi}$ and $\overline{\nabla^\mu J_\mu^{(i)}}$ for $i=4,5$ by removing these terms, so that we may write the scalar field equation \eqref{eq:PreHorndeskiScalarEom} as 
\begin{equation}
\left(\sum_{i=2,3} (\nabla^\mu J^{(i)}_\mu  -  P^{(i)}_\phi ) + \sum_{i=4,5} ( \overline{\nabla^\mu J_\mu^{(i)}} - \overline{P^{(i)}_\phi} )\right) - R \Xi  =0 \,,
\end{equation}
where $\Xi \equiv  G_{4\phi} + (G_{4X} - G_{5 \phi} )  \Box \phi - \frac{1}{2} G_{5X}(\Box \phi)^2$.
Inserting the trace of eq.~\eqref{eq:HorndeskiMetricEq}, we rewrite the scalar field equation as 
\begin{equation}
\label{eq:HorndeskiScalarEq}
\left(\sum_{i=2,3} (\nabla^\mu J^{(i)}_\mu  -  P^{(i)}_\phi ) + \sum_{i=4,5} ( \overline{\nabla^\mu J_\mu^{(i)}} - \overline{P^{(i)}_\phi} )\right) \Gamma + \Xi \sum_{i=2}^5 R^{(i)} = -\frac{T}{M_p^2} \Xi \,.
\end{equation}
The equations of motion \eqref{eq:HorndeskiMetricEq} and \eqref{eq:HorndeskiScalarEq} are now in the form that we will use in the rest of the paper.

There is an important pattern to notice in $R_{\mu\nu}^{(i)}$, $P_\phi^{(i)}$ and $\nabla^\mu J_\mu^{(i)}$ relating the functional derivatives of $G_i$ to their pre-factors of $\partial^s \phi$ for integers $s$.
For any term within one of these objects let $m$ be the number of functional derivatives with respect to $\phi$
acting on the $G_i$ contained in it.
Correspondingly, let $n$ be the number of functional derivatives with respect to $X$.
Consider the example
\begin{equation}
-\frac{1}{2}G_{2X} \nabla_\mu \phi \nabla_\nu \phi \in \mathcal{G}^{(2)}_{\mu\nu}
\end{equation}
with $n=1$ and $m=0$.
We can see that $2n+m=2$ derivatives of $\phi$ appear multiplying $G_{2X}$.
We find that this relation holds for all terms in $R^{(2)}_{\mu\nu}$.
Furthermore, we can examine all of the $R_{\mu\nu}^{(i)}$, $P^{(i)}_\phi$ and $\nabla^\mu J_\mu^{(i)}$ and find the relations listed in table~\ref{tab:Pattern}.
These observations will become very useful when studying the Einstein gravity limits of Horndeski theory in section~\ref{sec:GRHorndeski}.

\begin{table}[tbp]
\centering
\begin{tabular}{|l|c|c|c|}
\hline
 & 2n+m+1 & 2n+m  & 2n+m-1\\
\hline
i even & - & $R^{(i)}_{\mu\nu}$   & $P^{(i)}_\phi$, $\nabla^\mu J_\mu^{(i)}$ \\
i odd  & $R^{(i)}_{\mu\nu}$ & $P^{(i)}_\phi$, $\nabla^\mu J_\mu^{(i)}$  & - \\
\hline
\end{tabular}
\caption{\label{tab:Pattern} The number of factors of the form $\partial^s \phi$ that multiply $G_i$ within the functions $R^{(i)}$, $\nabla^\mu J^{(i)}_\mu$ and $P^{(i)}$, where $m$ is the number of functional derivatives with respect to $\phi$ and $n$ with respect to $X$ that act upon the $G_i$.}
\end{table}

Finally, we note that a cosmological propagation speed of gravitational waves at the speed of light places very tight constraints on non-vanishing contributions of $G_{4X}$ and non-constant $G_5$ and that with the direct detection of tensor waves~\citep{ObservationofGravitationalWavesfromaBinaryBlackHoleMerger} a corresponding measurement may soon be realized.

\section{Scaling to describe nonlinear regimes}
\label{sec:MyMethod}

In order to reproduce GR in the Solar System, where it has been well tested~\citep{TheConfrontationBetweenGeneralRelativityAndExperiment}, a theory of modified gravity that significantly deviates at large scales requires a screening mechanism that suppresses deviations locally.
Such screening effects depend on nonlinear terms in the gravitational equations of motion which, when dominant, prevent large deviations from GR.
Well known mechanisms include the Vainshtein~\citep{ToTheProblemOfNonvanishingGravitationMass}, chameleon \citep{ChameleonFieldsAwaitingSurprisesForTestsOfGravityInSpace} and k-mouflage \citep{KMOUFLAGEGravity} effects.
Theories employing these mechanisms usually introduce different regimes such that screening is active and GR is recovered near massive bodies or in high-density environments; whereas at large scales, there is no screening effect and gravity is modified.

Due to the reliance on nonlinear contributions, a linearisation by a low-energy, static perturbative expansion of the equations of motion cannot be applied to the screened regime, where the nonlinear terms dominate, and is only valid in a regime where the theory is not screened.
Hence, in the presence of screening one cannot use this expansion to test gravity in the local universe without solving the full equations of motion.
In the case of the Vainshtein mechanism, the perturbative expansion breaks down at the radius where the nonlinear terms start to dominate, i.e., when this characteristic screening radius is approached from the outside.
In chameleon models, the nonlinear potential cannot be linearised as it needs to cover a wide range of values to allow for the large increase of effective mass in high densities required for screening.

Hence, to find a perturbative expansion in the nonlinear regime for galileon gravity models, where Vainshtein screening operates, Refs.~\citep{ClassicalDualsOfDerivativelySelfCoupledTheories, ClassicalDualsLegendreTransformsAndTheVainshteinMechanism} proposed the use of dual Lagrangians obtained from Laplace transforms or Lagrange multipliers.
A dual Lagrangian is physically equivalent to the original Lagrangian but written in terms of auxiliary fields.
The benefit of this dual description is that when a low-energy, static expansion is performed, the natural regime which the expansion describes is within the nonlinear, and hence screened, regime of the original Lagrangian.
This expansion for the dual breaks down as one approaches a regime where the linear terms of the original Lagrangian dominate.
The expansion of the dual is therefore complimentary to the expansion of the original Lagrangian.
As a result, these dual methods allow for a comparison between the predictions of modified gravity theories in screened regions and observables in the local universe.
Ref.~\citep{TheParametrizedPostNewtonianVainshteinianFormalism} demonstrates how the Lagrange multiplier method can be used to perform a parametrised post-Newtonian expansion~\cite{TheoryAndExperimentInGravitationalPhysics} for derivatively coupled theories and gives an example of the expansion to second order for the cubic galileon model in the Jordan frame.
However, the dual methods become increasingly involved when applied to more complex Horndeski models and the transform is not always obvious.
A more concise method, enabling an expansion in the nonlinear, screened regime, should therefore be very useful to facilitate the analysis of deviations from GR in the local region.

In section~\ref{sec:Method}, we present a new, simpler method enabling this expansion that is based on the scaling of the scalar field within the metric equations of motion and scalar field equation and reproduces the known results from the dual approach of the cubic galileon.
We first demonstrate its operability with the explicit example of the cubic galilieon coupled to gravity in section~\ref{sec:ScalingDerivative}.
We then also show how this method applies to screening through a scalar field potential by examining the chameleon model in section~\ref{sec:Cham}, which has eluded these dual methods.

\subsection{A scaling method}
\label{sec:Method}

Let us first consider the heuristic form of a scalar field equation
\begin{equation}
\label{eq:testFn}
\alpha^s F_1(\phi,X) + \alpha^t F_2(\phi,X) = T/M_p^2
\end{equation}
for free functions $F_{1,2}$, the trace of the stress-energy tensor $T$ of a given matter distribution, and arbitrary real numbers $s$ and $t$.
Let $\alpha$ be an arbitrary coupling constant that controls the scale at which the different terms become important. 
Now, consider the expansion of the scalar field
\begin{equation}
\label{eq:phiExp}
\phi = \phi_0(1+\alpha^q \psi) \,,
\end{equation}
where we have separated out a constant part $\phi_0$ and a varying part $\psi$; $q$ is a real number which is determined by the gravitational model under consideration.
If $\alpha$ is large in comparison to $\psi$, then clearly for a sensible expansion, $q$ should be negative, whereas if $\alpha$ is small, $q\geq0$.
This ensures that the second term will be small provided $\psi < \alpha^{-q}$. We will see that the values that $q$ can take are restricted, and this will be the crux of the scaling method we propose here.

Let $F_{1,2}(\phi, X)$ scale homogeneously in $\alpha^q$ with respect to the expansion \eqref{eq:phiExp}, so that we get 
\begin{equation}
\label{eq:testFnAlpha}
\alpha^{s+mq} F_1(\psi,(\partial\psi)^2) + \alpha^{t+nq} F_2(\psi,(\partial\psi)^2) = T/M_p^2 \,,
\end{equation}
for real numbers $m$, $n$.
We now have the original equation \eqref{eq:testFn} cast as a function of $q$. 
Equation~\eqref{eq:testFnAlpha} needs to hold for arbitrary $\alpha$ but the right-hand side is not a function of $\alpha$, which implies that there must be a term on the left-hand side that is not a function of $\alpha$ either.
As a result, $q$ can only take on certain values, namely
\begin{equation}
\label{eq:TestSetOfQ}
q \in \left\lbrace -\frac{s}{m} , -\frac{t}{n} \right\rbrace \,.
\end{equation}

Next we wish to examine the case where $\alpha \gg \psi$ or $\alpha \ll \psi$, and so we take the formal limits of $\alpha \to \infty$ or $\alpha \to 0$, respectively.
It should be stressed that the physical value for such constants are given when one writes down a specific action, and that being constants, such limits do not involve changing the value for $\alpha$, rather the scale of $\alpha$ changes with respect to $\psi$.
In order for these limits to be meaningful, we need to ensure that no terms in eq.~\eqref{eq:testFnAlpha} diverge. For simplicity, we let $-\frac{s}{m} > 0 > -\frac{t}{n}$. So if we consider the case when $\alpha \to \infty$, we need to take the smaller of the two values for $q$ such that all powers of $\alpha$ that appear in eq.~\eqref{eq:testFnAlpha} are less than or equal to zero, preventing any divergences. This leads to the equations
\begin{align}
\alpha^{s+m(-\frac{t}{n})} F_1(\psi,(\partial\psi)^2) +  F_2(\psi,(\partial\psi)^2) & = T/M_p^2 \,, \nonumber \\
\alpha\rightarrow\infty : \ \ F_2(\psi,(\partial\psi)^2) & = T/M_p^2 \,.
\end{align}
Conversely, if we let $\alpha \to 0$, then all powers must be positive. Hence, we must take the largest possible value for $q$. In this case we get
\begin{align}
F_1(\psi,(\partial\psi)^2) + \alpha^{t+n(-\frac{s}{m})} F_2(\psi,(\partial\psi)^2) & = T/M_p^2 \,, \nonumber \\
\alpha \to 0 : \ \ F_1(\psi,(\partial\psi)^2) & = T/M_p^2 \,.
\end{align}
Hence, we have found two different equations of motion governing the dynamics of $\psi$ in the two different limits.
This allows us to perform a simplified perturbative expansion in each of the two limits which is valid in one region but not in the other and breaks down near $\psi \approx \alpha$, where one may want to impose some matching condition.
One can easily see that we can continue to add additional functions to eq.~\eqref{eq:testFn} to extract the dominating terms in the different limits. 

Besides the scalar field equation, we also need the equation of motion for the metric to be consistent in these limits.
We apply the same expansion in eq.~\eqref{eq:phiExp} to rewrite the metric field equation as a function of $g_{\mu\nu}$ and $\psi$, making it dependent on $q$.
Upon taking a limit of $\alpha$, the value of $q$ used in the metric field equation must be the same as that in the scalar field equation as it describes the same field.
For a sensible limit, the metric field equation should not diverge in the same limiting process described above.

The prescription given here applies broadly to the equations of motion that appear in different gravity theories.
There are, however, two special scenarios that we have to consider in more detail: (i)~one term is not a function of $\alpha$ after the expansion in eq.~\eqref{eq:phiExp}; and (ii)~the power of $\alpha$ is not a function of $q$.
We have insisted that in the limit of concern, $q$ must take a value that ensures a non-vanishing contribution to the left-hand side of the equation of motion \eqref{eq:testFnAlpha}.
In the first case, the term independent of $\alpha$ already provides a non-vanishing term in both limits, so we are left with the condition that no terms diverge.
Hence, the set of feasible values of $q$ in either limit creates an inequality condition for $q$, which requires that $q$ is equal to or less (greater) than the minimum (maximum) of the analogous set to eq.\eqref{eq:TestSetOfQ} for $\alpha\to \infty (0)$; we are then free to pick a value that satisfies this condition.
However, the consideration of further equations of motion may still provide a requirement for an exact value of $q$. 
The second scenario poses a problem when taking either one or the other limits of $\alpha$.
If for instance, we have a contribution of the form $\alpha^m$ with $m>0$, then this term will diverge in the limit of $\alpha \to \infty$ regardless of the value of $q$, but not for $\alpha \to 0$, and viceversa for $m<0$.
We can then only solve for $\psi$ in the limit where there are no divergences.

The extension of the scaling method to a full, higher-order expansion $\phi = \phi_0(1+\sum\alpha^i \psi_i)$ will be presented in separate work. 
In summary, the prescription for the first-order scaling approach is as follows: 
\begin{itemize}
  \item[(i)] expand the scalar field in the equations of motion according to eq.\eqref{eq:phiExp}: $\phi = \phi_0(1+\alpha^q \psi)$;
  \item[(ii)] for a field equation, find all values of $q$ where an exponent of $\alpha$ becomes 0;
  \item[(iii)] if taking the limit $\alpha \to 0$, $q$ takes at least the maximum value of this set;
  \item[(iv)] if taking the limit $\alpha \to \infty$, $q$ takes at most the minimum value of this set;
  \item[(v)] check that the complementary field equations do not diverge with this limit and value of $q$;
  \item[(vi)] should no terms diverge and should at least one non-vanishing term exist in all field equations for this $q$, the resulting equations of motion describe the fields in the corresponding limit.
\end{itemize}
In order to demonstrate the applicability of the scaling method introduced here for the description of the different screening mechanisms operating in Horndeski theory (see section~\ref{sec:HornGravity}), we start by providing two simple examples:
we first discuss the application of the scaling method to the derivative screening of the cubic galileon model in section~\ref{sec:ScalingDerivative} and then apply it to the scalar field screening in the chameleon model in section~\ref{sec:Cham}.
In section~\ref{sec:GRHorndeski}, we then discuss its application to the full Horndeski theory.

\subsection{Scaling with derivative screening}
\label{sec:ScalingDerivative}

Derivative terms in a field theory can be approximated by the energy of the system, which in the low-energy limit is generally smaller than their coefficients.
Thus, these contributions are generally suppressed and one would not expect terms involving derivatives other than the kinetic term to be relevant for low-energy physics.
However, it has been found that derivative terms can give rise to screening mechanisms, caused either by powers of $\partial \phi$ as in kinetic screening like k-mouflage, or by higher-derivative terms of the form $\partial ^2 \phi$ as in the Vainshtein mechanism.
Second-derivative terms in the Lagrangian are na\"{i}vely indicative of an Ostrogradsky ghost as one would expect the equations of motion to contain third time derivatives. 
However, it is possible to construct theories that despite containing higher derivatives, remain second order at the level of the equations of motion, i.e., galileon models or the generalised galileon, which are respectively Horndeski models.
These models are interesting as they arise as effective scalar-tensor theories in the decoupling limit of braneworld models~\citep{4DGravityOnABraneIn5DMinkowskiSpace} as well as in massive gravity~\citep{ResummationofMassiveGravity}.

As a demonstration of the scaling method introduced in section~\ref{sec:Method}, we now apply it to the cubic galileon model, which employs the Vainshtein screening mechanism.
The model is defined by the action
\begin{equation}
\label{eq:CubicAction}
S_\emph{cubic} = \frac{M_p^2}{2} \int d^4x \sqrt{-g} \left[ \phi R + \frac{2 \omega}{\phi} X - \frac{\alpha}{4}\frac{X}{\phi^3}\Box\phi \right] + S_m[g] \,,
\end{equation}
where $S_m$ denotes the minimally coupled matter action, $\omega$ is the Brans-Dicke parameter and $\alpha$ is the coupling strength with units $mass^{-2}$.
Note that the model is embedded in the Horndeski action, eq.~\eqref{eq:HornAction}, which can easily be seen by setting $G_2=	2\omega \phi^{-1} X$, $G_3 = \alpha \phi^{-3}X/4$, $G_4 = \phi$, and $G_5 = 0$.
With this choice of the $G_i$ functions, the metric field equations become 
\begin{equation}
\label{eq:CubicMetricEom}
\phi R_{\mu \nu} = M_p^{-2} \left[T_{\mu \nu} - \frac{1}{2} T g_{\mu \nu} \right] + \frac{\omega}{\phi} \nabla_\mu \phi \nabla_\nu \phi+ \frac{1}{2} \Box \phi g_{\mu \nu}+\nabla_{\mu} \nabla_\nu \phi
+ \frac{\alpha}{8} \left[ \phi^{-3} \mathcal{M}^{(3)}_{\mu\nu} + \phi^{-4} \mathcal{M}^{(4)}_{\mu\nu} \right] ,
\end{equation}
where we have introduced the rank-2 tensors 
\begin{align*}
\mathcal{M}^{(3)}_{\mu\nu} &\equiv - X \Box \phi g_{\mu \nu} - \Box \phi \nabla_\mu \phi \nabla_\nu \phi - \nabla_\mu X \nabla_\nu \phi- \nabla_\mu \phi \nabla_\nu X \,,
\\
\mathcal{M}^{(4)}_{\mu\nu} &\equiv 6 X \nabla_\mu \phi \nabla_\nu \phi \,.
\end{align*}
The scalar field equation becomes
\begin{equation}
\label{eq:CubicScalarEom}
(3+2\omega) \Box \phi + \frac{\alpha}{4}\left[ \phi^{-2} \mathcal{S}^{(2)} + \phi^{-3} \mathcal{S}^{(3)} + \phi^{-4} \mathcal{S}^{(4)} \right] = M_p^{-2} T \,,
\end{equation}
where we have defined the scalar quantities
\begin{align*}
\mathcal{S}^{(2)} &\equiv - (\Box \phi)^2 - \nabla^\mu \phi \nabla_\mu \Box \phi - \Box X \,,
\\
\mathcal{S}^{(3)} &\equiv  5\nabla_\mu \phi \nabla^\mu X - X\Box \phi  \,,
\\
\mathcal{S}^{(4)} &\equiv 18 X^2 \,.
\end{align*}
The introduction of $\mathcal{M}$ and $\mathcal{S}$ facilitates the analysis of the equations of motion; the superscripts describe the power to which the scalar field appears inside of them, so that $\mathcal{M}$ and $\mathcal{S}$ are homogeneous polynomials with respect to $\partial^n\phi$ for $n=1,2$.

As we want to describe the cases where the interaction terms are dominant or vanishing, the scaling parameter in eq.~\eqref{eq:phiExp} is given by the coupling strength $\alpha$. 
This makes the $\mathcal{S}^{(i)}$ and $\mathcal{M}^{(i)}_{\mu\nu}$ become functions of $\psi$ that scale homogeneously with respect to $\alpha^q$ with degree of their superscript.
Performing the expansion, terms on the left-hand side of eq.~\eqref{eq:CubicScalarEom} become functions of $\alpha$, while the right-hand side remains a function of the stress-energy tensor only.
From eq.~\eqref{eq:CubicScalarEom} one can easily identify the exponents of $\alpha$ that appear in the expansion.
For example, $\alpha \mathcal{S}^{(2)}(\phi) \to \alpha^{1+2q} \phi_0^2 \mathcal{S}^{(2)}(\psi)$ gives the exponent $1+2q$. 
The remaining exponents from $\mathcal{S}$ are $q$, $1+3q$, and $1+4q$ . 
In the limits of large or small $\alpha$, one term on the left-hand side of eq.~\eqref{eq:CubicScalarEom} must balance the right-hand side by being independent of $\alpha$.
This puts restrictions on the values of $q$ as at least one of the exponents must be zero, namely,
\begin{equation}
\label{eq:CubicAllowedQ1}
q \in \left\lbrace  0, -\frac{1}{4}, -\frac{1}{3}, -\frac{1}{2} \right\rbrace = Q_\mathcal{S} \,.
\end{equation}
We expect that the limit of $\alpha \to \infty$ corresponds to the limit where screening dominates and hence where Einstein gravity is recovered.
To prevent any terms from diverging in this regime, all powers of $\alpha$ in eq. \eqref{eq:CubicScalarEom} should be less than or equal to zero.
This requirement and the restriction that $q\in Q_\mathcal{S}$ implies that we must adopt the smallest value in $Q_\mathcal{S}$, $q = \min(Q_\mathcal{S})=-\frac{1}{2}$.

Next, we must also find the corresponding exponents of $\alpha$ in the metric field equation, which are 
\begin{equation}
\label{eq:CubicAllowedQ2}
\left\lbrace  0, -\frac{1}{4}, -\frac{1}{3} \right\rbrace = Q_\mathcal{M} \,.
\end{equation}
However, eq.~\eqref{eq:CubicMetricEom} contains the term $\phi_0 R_{\mu\nu}$ which is not a function of $\alpha$, and so we do not require that the value of $q$ must be in $Q_\mathcal{M}$.
The condition that no terms diverge implies that $q\leq \min (Q_\mathcal{M}) = -\frac{1}{3}$. Hence the value $q=-\frac{1}{2}$ is allowed by both field equations.

Adopting this value for $q$ and taking the limit of large $\alpha$, eq.~\eqref{eq:CubicMetricEom} and \eqref{eq:CubicScalarEom} become 
\begin{equation}
\label{eq:CubicMetricEomInfty}
\phi_0 R_{\mu \nu} = M_p^{-2} \left[T_{\mu \nu} - \frac{1}{2} T g_{\mu \nu} \right] \,,
\end{equation}
\begin{equation}
\label{eq:CubicScalarEomInfty}
\frac{1}{4} \mathcal{S}^{(2)} (\psi)   = M_p^{-2} T \,,
\end{equation}
where the galileon field scales as 
\begin{equation}
\phi = \phi_0(1+\alpha^{-1/2} \psi) \,.
\end{equation}
One can see that the metric field equation has reduced to Einstein's field equations (when setting $\phi_0=1$), indicating a screening effect.
Hence, we have recovered the known result of Vainshtein screening in cubic galileon gravity when the self-interaction term dominates, and we have also found the scalar field equation that $\psi$ and thus $\phi$ satisfies in this limit. 

In contrast, for the opposite limit of $\alpha \to 0$, we expect no screening effect.
In order to prevent divergences in this limit, the powers of $\alpha$ must be greater than or equal to zero, and thus $q = \max(Q_\mathcal{S})=0$.
The field equations then become 
\begin{equation}
\label{eq:CubicMetricEomZero}
\phi R_{\mu \nu} = M_p^{-2} \left[T_{\mu \nu} - \frac{1}{2} T g_{\mu \nu} \right] + \frac{\phi_0^2}{\phi} \omega \nabla_\mu \psi \nabla_\nu \psi+ \frac{\phi_0}{2} \Box \psi g_{\mu \nu} + \phi_0 \nabla_{\mu} \nabla_\nu \psi \,,
\end{equation}
\begin{equation}
\label{eq:CubicScalarEomZero}
\phi_0 \Box \psi   = M_p^{-2} T \,,
\end{equation}
and the galileon field scales as 
\begin{equation}
\phi= \phi_0(1+\psi) \,.
\end{equation}
We recognise these relations as the equations of motion of Brans-Dicke gravity in the Jordan frame~\citep{MachsPrincipleAndARelativisticTheoryOfGravitation}.
Hence, we have recovered the metric and scalar field equations describing the cubic galileon model in the deeply screened and unscreened limits.
All steps taken in the process were trivial, demonstrating the simplicity and efficiency of our scaling method.

Strictly speaking, one must also Taylor-expand the negative powers of $\phi$ that appear in the equation of motion. 
This contributes extra values of $q$ into the sets~(\ref{eq:CubicAllowedQ1}) and (\ref{eq:CubicAllowedQ2}). 
However, these will come from terms $\alpha^{a+bq+iq}$ with positive integers $i$, and for simplicity $a,b>0$, contributing extra $q$ values of $-a/(b+i)$.
The minimum of these additional values is assumed when $i=0$.
Hence, adopting this minimum corresponds to only taking the first, constant, term in a Taylor expansion of the negative powers of $\phi$, i.e., the power of $\phi_0$, and neglecting higher orders.
The other relevant value is the maximum, which is $q\rightarrow0$ when $i \to \infty$ and is already included in the sets.
We have omitted these terms for simplicity in the calculations above.

\subsection{Scaling with local field value screening}
\label{sec:Cham}

Besides adding powers of derivatives to an action, one may add a self-interaction potential such as a mass term.
In applying the expansion in eq.~\eqref{eq:phiExp}, we have so far relied on derivatives such that the constant part vanishes and $\psi$ is separated out with a factor of $\alpha$ to some power in each term.
This allowed the direct manipulation of the equations of motion.
However, screening with a scalar field potential relies upon the field assuming a particular value that minimises an effective potential of scalar field and matter density. In doing so, the scalar field acquires an effective mass that is dependent of the ambient mass density.
We shall consider here the chameleon mechanism, which operates by forcing the field to become more massive in high-density regions, and so satisfies stringent Solar System tests.

To demonstrate how chameleon screening emerges in our framework, we consider the action
\begin{equation}
\label{eq:ChamAction}
S_\emph{cham}=\frac{M_p^2}{2}\int d^4x \sqrt{-g}\left[ \phi R + \frac{2 \omega}{\phi} X - \alpha(\phi-\phi_\emph{min})^n \right] + S_m[g] \,,
\end{equation}
where $n$ is a constant, $\alpha$ describes a coupling constant, and $\phi_\emph{min}$ denotes the value that minimises the potential (for $n>0$).
The action can be embedded in Horndeski theory by setting $G_2 = - 2 \omega\phi^{-1} X - \alpha(\phi-\phi_\emph{min})^n $, $G_3 = 0$, $G_4 = \phi$, and $G_5 =0$.
The equations of motion are 
\begin{align}
\label{eq:ChamScalarEom}(3+2\omega)\Box \phi =& M_p^{-2} T + \alpha (\phi - \phi_{min})^{n-1} (2(\phi- \phi_{min}) - n \phi) \equiv V'_{eff}(\phi) \,, \\
\label{eq:ChamMetricEom}\phi R_{\mu\nu} =& M_p^{-2} [T_{\mu \nu}- \frac{1}{2} g_{\mu\nu}T] +\frac{\omega}{\phi}\nabla_\mu \phi \nabla_\nu \phi +\frac{1}{2}\Box \phi g_{\mu\nu} \nonumber \\
&+\nabla_\mu\nabla_\nu\phi -\frac{1}{2}g_{\mu\nu}\alpha(\phi-\phi_{min})^n \,.
\end{align}
We again use the coupling constant in the expansion of $\phi$.
However the potential provides powers of $\alpha$ that are dependent on the value of $\phi_0$.
We consider two cases: (i) $\phi_0 \approx \phi_\emph{min}$; and (ii) $\lvert\phi_0-\phi_\emph{min}\rvert \gg \phi_0 \alpha^q\psi$.

For (i), $\phi_0 - \phi_\emph{min} \approx 0$ approximately minimises the self-interaction potential.  This results in $\alpha^q \psi$ becoming the argument of the potential terms and thus dependent on $q$. We
obtain the factors $\alpha^{1+(n-1)q}$ and $\alpha^{1+nq}$ in the scalar field equation~\eqref{eq:ChamScalarEom}, which together with the powers from the kinetic term implies that 
\begin{equation}
\label{eq:qInCham}
q \in \left\lbrace 0, -\frac{1}{n}, \frac{1}{1-n} \right\rbrace = Q_{\mathcal{S}} \,.
\end{equation}
The maximum and minimum values in $Q_{\mathcal{S}}$ thus depend on the value of $n$.
For $q=0$ and $\alpha\to0$, the scalar field equation describes a free scalar field sourced by the trace of the stress-energy tensor regardless of the value of $n$.

Consider the case when $q= \frac{1}{1-n}>0$ and $\alpha\to 0$; the field is no longer dynamical as no gradients appear in the scalar field equation.
Rearranging for $\psi$ gives the value which minimises $V_\emph{eff}$ in this limit,
\begin{equation}
\label{eq:ChamSol}
\psi=\left( -\frac{M_p^{-2} T }{ n \, \phi_0^n}\right)^\frac{1}{n-1} \,.
\end{equation}
We observe that for $n<1$, $\psi$ is suppressed for large $|T|$ but relevant when $|T|$ is small, which is as expected for the chameleon screening mechanism. 
However, note that $n<1$ represents the limit of $\alpha \to 0$, not $\infty$ as in the screened limit for the Vainshtein mechanism (section~\ref{sec:ScalingDerivative}).

In the metric field equation~\eqref{eq:ChamMetricEom}, derivatives contribute to the powers of $\alpha$ as $q$ and $2q$ whereas the potential carries the power $1+nq$. 
Upon examination, one finds that the $n<1$ chameleon case requires an additional restriction.
In order for the metric field equation to not diverge in the limit of $\alpha\to 0$ with $q=\frac{1}{1-n}>0$, the term proportional to $\alpha^{1+n/(1-n)}$ must have a positive exponent.
This is true for $n>0$, reproducing that for chameleon screening the potential must have an exponent $0<n<1$ (see, e.g.,~\citep{HaloModellingInChameleonTheories}). 
One can also see this from $Q_{\mathcal{S}}$ as we examined the case when $\frac{1}{1-n}> -\frac{1}{n}$, which requires that $n\in (0,1)$.

For (ii), the potential is a power of $( \Delta \phi + \phi_0 \alpha^q \psi )$ with $\Delta \phi \equiv \phi_0 - \phi_\emph{min}$.
Here, $\Delta\phi$ is the dominant term as by definition $\phi_0 \alpha^q \psi$ is a small perturbation, so a Taylor expansion of the potential in $\psi$ then gives
\begin{equation}
\alpha ( (\Delta \phi)^n  +n (\Delta \phi)^{n-1} \alpha^q \phi_0 \psi + ... ) \,.
\end{equation}
From this we see that the relevant set of values that $q$ must take for the equation of motion \eqref{eq:ChamScalarEom} to have a term that goes as $\alpha^0$ is 
\begin{equation}
\left\lbrace 0,-\frac{1}{i} \right\rbrace
\label{eq:chameleonsettwo}
\end{equation}
for positive integers $i$. 

However, we now have a power of $\alpha$ that is not a function of $q$. As such, we cannot take the $\alpha\to \infty$ limit without causing a divergence in the field equations. We are left to consider the limit of $\alpha \to 0$, where the maximum of the set in eq.~(\ref{eq:chameleonsettwo}) implies $q=0$. In this limit the scalar field equation becomes one of a free field sourced by the stress-energy tensor and the metric field equation becomes that of Brans-Dicke theory.

\section{Einstein limits in Horndeski gravity}
\label{sec:GRHorndeski}

A viable theory of gravity must have the capability of reducing to Einstein gravity in the Solar System, so the Horndeski-type actions of interest must provide such a limit.
In this section, we describe a novel method that allows an efficient assessment of whether an action can assume an Einstein limit or not.
This is achieved by examining the powers of $\alpha$, our limiting parameter, that appear in the action and employ the scaling procedure introduced in section~\ref{sec:MyMethod}.
Insisting that the metric field equations become Einstein's equations in a given limit and that the related scalar field equation does not diverge amounts to a set of two inequalities on the value of $q$, the exponent of $\alpha$ in the expansion~\eqref{eq:phiExp}.
To find these conditions, we examine the form of the equations of motion in Horndeski gravity written in terms of $\alpha^q \psi$.
Recurring patterns in these equations, identified in table~\ref{tab:Pattern}, allow a construction of the sets of all powers of $\alpha$ that the field equations contain.
The inequalities check for consistency between the extrema of these sets and allow us to determine whether the gravity theory of concern possesses a limit where the Einstein field equations are recovered.

We emphasise, however, that the consistency of the Einstein limit alone does not guarantee that Einstein gravity is recovered due to the operation of a screening effect.
We demonstrate this with examples of known screened and non-screened gravity theories which possess an Einstein gravity limit.
As we will show, one can, however, assess whether the recovery of Einstein gravity can be attributed to a screening effect or not by assuming a radial profile of the scalar field and examining the range of validity of the limit adopted.

In section~\ref{sec:Expansion} we outline the expansion of the Horndeski functions $G_i$ adopted and set up the tools needed to identify the embedded Einstein limits.
Section~\ref{sec:ScreeningConditions} focuses on finding all powers of $\alpha$ that can appear in the field equations given the free $G_i$ and hence determine the conditions on $q$.
We then use these conditions in section~\ref{sec:Examples} to find the limits of Einstein gravity.
Finally, we close with the discussion on how to assess whether these can be attributed to a screening effect in section~\ref{sec:RadialDependence}.

Note that while our analysis is performed on the Horndeski action, one could also extend it to beyond-Horndeski theories \citep{TransformingGravityTheoriesBeyondTheHorndeskiLagrangian,HealthyTheoriesBeyondHorndeski}.

\subsection{Expansion of $G_i$ functions and implications}
\label{sec:Expansion}

In order to apply the scaling method to Horndeski gravity, we first need to find a sensible description of the four generic $G_i$ function in the action \eqref{eq:HornAction}.
We adopt an expansion of the form
\begin{equation}
\label{eq:GenericF}
F(\phi, X) = \sum_{(m,n)\in I} \alpha^{p_{mn}} M_p^{-2m} X^m \prod_{\phi_i \in \mathcal{P}_{mn}} (\phi - \phi_i)^{p_{\phi_i}} \,,
\end{equation}
which embeds the galileon, chameleon, and k-mouflage actions. 
Hereby, $I$ denotes a set of indices and the $\alpha^{p_{mn}}$ are coefficients that determine when the terms they multiply become important.
Further, $\mathcal{P}_{mn}$ are sets of constants $\phi_i$ which only appear at most once per set and $p_{\phi_i}$ indicates the corresponding exponent of the scalar field potential for this $\phi_i$.
Note that we do not consider different parameters to describe the couplings, e.g., a combination of $\alpha$ and $\beta$; this is because these parameters are constants with the particular relationship between them set by the action, hence, limits in these couplings are taken simultaneously.
Additional powers of $M_p$ are needed in the expansion of $G_i$ as unlike $F$, these are not unitless.

In our scaling method, we only need to consider the powers of $\alpha$ that appear in the expansion of $F$ with respect to eq.~\eqref{eq:phiExp}, so we define $\alpha[\cdot]$ to be the set of powers of $\alpha$ which prefactor all terms in the equations of motion after performing the expansion.
More specifically, we have
\begin{equation}
\alpha[ F ] = \bigcup_{(m,n)\in I} \bigg\lbrace \alpha^{p_{mn}+2mq + \sum_{\phi_i \in \mathcal{P}_{mn}}  p_{\phi_i}q\delta(\phi_0 - \phi_i)} \bigg\rbrace \,,
\end{equation}
where $\delta$ denotes the Kronecker $\delta$-function, evaluating to unity when the argument is zero and vanishing otherwise.
In addition, several functional derivatives of $F$ need to be known to describe the equations of motion.
Again we only need to consider the powers of $\alpha$ that can appear.
Applying $\alpha[\cdot]$, we get 
\begin{align}
\alpha[F_X] &= \bigcup_{(m,n)\in I} \bigg\lbrace m \alpha^{p_{mn}+2(m-1)q + \sum_{\phi_i \in \mathcal{P}_{mn}}  p_{\phi_i}q\delta(\phi_0 - \phi_i)}  \bigg\rbrace \,, \\
\alpha[F_\phi ]&= \bigcup_{(m,n)\in I} \bigcup_{\phi_i \in \mathcal{P}_{mn}} \bigg\lbrace  p_{\phi_i} \alpha^{p_{mn}+2mq + \sum_{\phi_j \in \mathcal{P}_{mn}} q(p_{\phi_j}-\delta(\phi_j - \phi_i))\delta(\phi_0 - \phi_j)} \bigg\rbrace \,, \\
\alpha[F_{XX} ]&= \bigcup_{(m,n)\in I} \bigg\lbrace m(m-1)\alpha^{p_{mn}+2(m-2)q +  \sum_{\phi_i \in \mathcal{P}_{mn}} p_{\phi_i}q\delta(\phi_0 - \phi_i)} \bigg\rbrace \,, \\
\alpha[F_{X\phi}] &=  \bigcup_{(m,n)\in I} \bigcup_{\phi_i \in \mathcal{P}_{mn}} \bigg\lbrace m  p_{\phi_i} \alpha^{p_{mn}+2(m-1)q + \sum_{\phi_j \in \mathcal{P}_{mn}} q(p_{\phi_j}-\delta(\phi_j - \phi_i))\delta(\phi_0 - \phi_j)} \bigg\rbrace \,.
\end{align}
For our discussion, we shall define the multiplication of a set by $\alpha^s$ as the multiplication of all elements of the set by $\alpha^s$, which yields a new set.

To find an Einstein gravity limit for a general Horndeski theory, we apply $\alpha[\cdot]$ to the equations of motion to extract a set of values of $q$ a given gravity theory can assume to prevent divergences.
The minima and maxima of this set then determine whether the theory possesses an Einstein gravity limit.
To do this, we separate the equations of motion into suitable sub-components.
For instance, consider a single collection of terms from the metric field equation~\eqref{eq:HorndeskiMetricEq} with the simplest case of $\alpha[R^{(2)}_{\mu\nu}]$.
Using the expansion of $R^{(2)}_{\mu\nu}$ given in appendix~\ref{app:HornFieldEqs}, we find that the powers of $\alpha$ arise from combinations of $G_2$ and $G_{2X}$.
Using the relations in table~\ref{tab:Pattern} we can thus write
\begin{equation}
\label{eq:alphaInclusionExample}
\alpha[R^{(2)}_{\mu\nu}] \subset \alpha^{2q} \alpha[G_{2X}] \cup \alpha[G_2] = \alpha[G_2] \,.
\end{equation}
The reverse inclusion is the subject of the next section.
This shows that not all functional derivatives in the equations of motion need to be considered to determine $q$ since, in general, the values found from a functional derivative of $F$ are not independent of those found from $F$ itself.
We can summarise this conclusion as 
\begin{align}
\label{eq:GammaSubsets}
\alpha[F] &\supset \alpha^{2q} \alpha[F_X] \supset \alpha^{4q} \alpha[F_{XX}] \supset \alpha^{6q} \alpha[F_{XX}] \,, \\
\alpha[F_\phi] &\supset \alpha^{2q} \alpha[F_{X\phi}] \supset \alpha^{4q} \alpha[F_{XX\phi}] \,, \\
\alpha[F_{\phi\phi}] &\supset \alpha^{2q} \alpha[F_{X\phi\phi}] \,. \label{eq:GammaSubsetThree}
\end{align}
No further derivatives appear in the Horndeski equations of motion and are thus not required to determine $q$.
Hence, we only need to consider the functional derivatives which are highest on these chains.

Unlike with $\alpha^{2q}\alpha[F_X]$, we generally do not have $\alpha^q\alpha[F_\phi] \subset \alpha[F]$ as can be demonstrated with the counterexample $F=X^m (\phi-\phi_{min})^n$. 
The relevant set of powers then are $\alpha^q\alpha[F_\phi] = \{ \alpha^{(2m +1 + (n-1)\delta(\phi_0-\phi_{min})q}\}$ in one case and $\alpha[F] = \{ \alpha^{(2m +n\delta(\phi_0-\phi_{min}))q} \}$ in the other.
In the limit of $\phi_0=\phi_{min}$ these two sets are equivalent: $\{\alpha^{(2m+n)q}\}$.
However, when $\phi_0 \neq \phi_{min}$, the Kronecker $\delta$-functions vanish such that the two sets become $\alpha^q \alpha[F_\phi] = \{\alpha^{(2m+1)q}\}$ and $\alpha[F] = \{\alpha^{2qm}\}$, which differs in general.
This implies that we must have multiple inclusion chains for the functional derivatives with respect to $\phi$. 

\subsection{Screening conditions}
\label{sec:ScreeningConditions}

We now discuss the powers of $\alpha$ that appear in the Horndeski equations of motion.
We first focus on the metric field equation in section \ref{sec:ScreeningConMetric} and then on the scalar field equation in section \ref{sec:ScreeningConScalar}.
Next, we utilise the identification of those powers to directly infer a set of conditions on the four free function in the Horndeski action that they must satisfy in order to provide a limit where Einstein's field equations are recovered.
We summarise those in section~\ref{sec:EinsteinLimit}.
For clarity of the discussion, as there are two field equations, we define $q_\emph{metric}$ and $q_\emph{scalar}$ to be the values of $q$ dictated by the metric and scalar field equation, respectively. 
But because there is just one scalar field, we ultimately require that $q_\emph{metric} = q_\emph{scalar}$.

\subsubsection{Metric field equation}
\label{sec:ScreeningConMetric}
  
For the metric field equations~\eqref{eq:HorndeskiMetricEq} to be considered screened, we require Einstein's field equations to be recovered up to a rescaled, effective Planck mass, which should be obtained after applying the expansion~\eqref{eq:phiExp} and taking one of the limits in $\alpha$.
Examining eq.~\eqref{eq:HorndeskiMetricEq}, we see that this corresponds to the requirement that $\Gamma \to \epsilon$ and $\sum _{i=2}^5 R_{\mu\nu}^{(i)} \to 0$ for some constant~$\epsilon$.

Recall that $\Gamma = G_4 - X G_{5X}$.
There are two scenarios for which $\Gamma \to \epsilon$: (i) when $\Gamma$ contains a constant term with all other terms vanishing upon taking the limit; and (ii) when $\Gamma$ contains a potential term of the form $(\phi_\emph{min}-\phi)^n$ which is not minimised such that the term $(\phi_\emph{min}-\phi_0)^n$ remains in the equations after taking the limit.
Both scenarios recover Einstein's equations up to an effective Planck mass and can be expressed as
\begin{equation}
\label{eq:G4limitCondition}
 \Gamma \to \sum_n \alpha^0 \prod_{\phi_i \in \mathcal{P}_{m n}} (\phi_0 - \phi_{i})^{p_{\phi_i}} = \epsilon < \infty \,,
\end{equation}
where $n$ is an integer index and $p_{\phi_i}$ denotes an exponent in the potential. 
Note that this term is independent of $\alpha$ before the expansion \eqref{eq:phiExp} and hence is the leading term when performing the expansion.
We shall denote the value of the maximum or minimum $q$ found from $\alpha[\Gamma]$ as $q_\Gamma$. 
The requirement that only terms which scale as $\alpha^0$ do not vanish when taking a limit becomes an inequality on the value of $q_\emph{metric}$.
Choosing any value for $q_\emph{metric}$ beyond $q_\Gamma$ ensures that all terms which are functions of $q_\emph{metric}$ will go to zero, and as such we are only left with $\Gamma \to \epsilon$. Thus we must have $q_\emph{metric}<q_\Gamma$ for $\alpha\to \infty$ (or $q_\emph{metric}>q_\Gamma$ for $\alpha\to 0$).
The set of coefficients we need to consider to check these conditions is
\begin{equation}
\label{eq:AlphaGamma}
\alpha[\Gamma] = \alpha[G_4 - X G_{5X}].
\end{equation}
One must remove all terms which go as $\alpha^0$ from this set in order to find the minimum or the maximum value for $q_\Gamma$. 

The requirement that $\sum _{i=2}^5 R_{\mu\nu}^{(i)} \to 0$ removes the contribution of an effective stress-energy component attributed to the scalar field from the metric field equation.
In principle, one could also allow for this limit to tend to a cosmological constant, which, however, we assume to be negligible in regions where screening operates.
Further, this implies that we insist that no terms in this sum scales as $\alpha^0$ after our expansion.

Our aim is to use the chains of inclusion, eqs.~\eqref{eq:GammaSubsets}-\eqref{eq:GammaSubsetThree}, to reduce the number of terms we need to consider in the equations of motion \eqref{eq:HorndeskiMetricEq} and \eqref{eq:HorndeskiScalarEq}.
A collection of such chains exists for each of the free functions $G_i$. 
By examining the highest sets in those chains, we can identify all powers of $\alpha$ that appear in the metric and scalar field equations.
These highest sets will also contain degenerate terms, which further reduces the number of terms that we need to consider. 
By degenerate we mean that the resulting powers of $\alpha$ found from examining these particular terms are the same (see table~\ref{tab:Pattern}).

However, we are not interested in each $\alpha[G_i]$ individually, rather in the powers of $\alpha$ that arise in the full equations of motion. For the metric field equation specifically, we are interested in
\begin{equation}
 \mathcal{M}\equiv\alpha[ \, \sum_{i=2}^5 R_{\mu\nu}^{(i)} \, ] \,.
\end{equation}
The minimum or maximum values of $q$ in $\mathcal{M}$, which we shall denote as $q_\mathcal{M}$, will allow us to directly assess if the choice of gravitational action has an Einstein gravity limit or not.
In specific, we demand that the value of $q$ from the metric field equation satisfies $q_{metric}<q_\mathcal{M}$ in the $\alpha \to \infty$ limit (or $q_{metric}>q_\mathcal{M}$ when $\alpha \to 0$).

We now determine the relation between $\mathcal{M}$ and the highest elements of the chains in eq.~\eqref{eq:GammaSubsets}-\eqref{eq:GammaSubsetThree} so that our analysis can be simplified through the arguments outlined above.
This is non-trivial as while it is clear that $\mathcal{M}$ is included in the union of these sets, e.g., as in eq.~\eqref{eq:alphaInclusionExample}, terms from different $R^{(i)}_{\mu\nu}$ may mutually cancel out when taking the sum.
Hence, it is not a priori clear whether this inclusion can be reversed; all we know is that $\mathcal{M} \subset \bigcup_{i=2}^5\alpha[ \,  R_{\mu\nu}^{(i)} \, ]$.
Without the reversal, we cannot be sure that the powers of $\alpha^q$ we are using will remain in the sum, even though should they exist within any $R_{\mu\nu}^{(i)}$ individually.  

To proceed, we draw the analogy to a vector space over the free functions $G_i$ and their functional derivatives. 
The basis vectors are multiples of $\Box \phi$, $\nabla_\mu \nabla_\nu \phi$, the Ricci scalar, the Ricci tensor, and the Riemann tensor; $\sum_{i=2}^5 R_{\mu\nu}^{(i)}$ is then an element of this space.
Our method for showing equality between the sets is then to identify terms with coefficients that are highest in the chains of $G_i$ and linearly independent such that they cannot vanish unless they are already set to zero to begin with.
In doing so, any power of $\alpha$ that appears in a linearly independent term will not vanish in the sum. 
Finding such terms for all the highest sets in our chains of each $G_{i}$ will show that every power in the union appears in the sum, and therefore in $\mathcal{M}$, showing the reverse inclusion.

In the contribution of $R^{(5)}_{\mu\nu}$ to the metric field equations, one can see that the only terms with basis vectors that do not contribute to $R^{(i\neq5)}_{\mu\nu}$ are
\begin{align}
&G_{5X} R^{\alpha\beta} \nabla_\alpha \phi \nabla_\beta \nabla_{(\mu} \phi \nabla_{\nu)} \phi \,, \\
&G_{5\phi} R_{\alpha(\mu\nu)\beta} \nabla^\alpha \phi \nabla^\beta \phi \,,
\end{align}
which cannot vanish unless $G_{5X}=0$ or $G_{5\phi}=0$, respectively.
Thus we have that $\mathcal{M} \supset \alpha^{3q} \alpha[G_{5X}] \cup \alpha^{2q} \alpha[G_{5\phi}]$.
Similarly for $R^{(4)}_{\mu\nu}$, we find that the terms
\begin{align}
\label{eq:G4XAlpha} &G_{4X} R \nabla_\mu \phi \nabla_\nu \phi \,, \\
&G_{4\phi}  \nabla_\mu \nabla_\nu \phi
\end{align}
are linearly independent from the rest of the summands in $\sum R^{(i)}_{\mu\nu}$ such that the exponents of $\alpha$ are given by $\mathcal{M} \supset \alpha^q\alpha[G_{4\phi}] \cup \alpha^{2q}\alpha[G_{4X}]$.
In $R^{(3)}_{\mu\nu}$, we can isolate the term
\begin{equation}
\label{eq:G3XAlpha}
G_{3X} \nabla_{(\mu} X \nabla_{\nu)} \phi \,.
\end{equation}
Thus, the functions $G_{5X}$, $G_{5\phi}$, $G_{4X}$, $G_{4\phi}$, and $G_{3X}$ are all sole coefficient of a linearly-independent vector in the operator vector space.
However, the remaining terms in $R^{(i)}_{\mu\nu}$ do not appear alone in independent terms: $G_{2X}$, $G_2$, $G_{3\phi}$, $G_{4\phi\phi}$, and $G_{5\phi\phi}$ cannot be considered in isolation. Hence, the powers of $\alpha$ we would get by including them individually may not be present upon taking the sum due to possible mutual cancellations.

Let us first examine $G_{5\phi\phi}$ and consider, for instance, the terms proportional to (or in direction of) $\nabla_\mu \phi \nabla_\nu \phi \Box\phi$, which are
\begin{equation}
\label{eq:G5phiphiAlpha}
\nabla_\mu \phi \nabla_\nu \phi \Box\phi \bigg\lbrace \frac{1}{2} G_{5\phi\phi} -2 G_{4X\phi} +\frac{1}{2} G_{3X} \bigg \rbrace \,.
\end{equation}
If any term in $\frac{1}{2}G_{5\phi\phi}$ within the combination~\eqref{eq:G5phiphiAlpha} is cancelled, this implies that the same term must also appear in $-2G_{4X\phi}+\frac{1}{2}G_{3X}$.
As can be seen from eqs.~\eqref{eq:G4XAlpha} and \eqref{eq:G3XAlpha}, $\alpha[G_{4X\phi}]$ and $\alpha[G_{3X}]$ are contained in coefficients of independent terms found in $R^{(4)}_{\mu\nu}$ and $R^{(3)}_{\mu\nu}$, respectively.
Therefore, any term that could be cancelled in $\frac{1}{2}G_{5\phi\phi}$ must still contribute to $\mathcal{M}$ through these independent terms and we can simply include the whole set $\alpha[G_{5\phi\phi}]$ in $\mathcal{M}$. Thus we have that $\mathcal{M}\supset\alpha^{3q} \alpha[G_{5X}] \cup \alpha^{2q} \alpha[G_{5\phi}] \cup \alpha^q\alpha[G_{5\phi\phi}] \supset \alpha [R^{(5)}_{\mu\nu}]$.

Now consider the remaining non-independent functions in $R^{(i)}_{\mu\nu}$, i.e., $G_{2X}$, $G_2$, $G_{3\phi}$, and $G_{4\phi\phi}$.
There are two equations that mix the four:
\begin{align}
\label{eq:FindingAlphaK1}
&\nabla_\mu \phi \nabla_\nu \phi ( G_{3\phi} - \frac{1}{2} G_{2X} - G_{4\phi\phi}) \,, \\
\label{eq:FindingAlphaK2}
&-\frac{1}{2}g_{\mu\nu}( X G_{2X}  - G_2 - 2X G_{4\phi\phi}) \,.
\end{align}
Hence, it is possible that terms in $G_{2X}$, $G_2$, $G_{3\phi}$, and $G_{4\phi\phi}$ cancel in their contribution to $R^{(i)}_{\mu\nu}$, which if they do should be excluded when determining $q$.
For the general scenario and for simplicity, we therefore impose that no terms contained in the four functions $G_{2X}$, $G_2$, $G_{3\phi}$, and $G_{4\phi\phi}$, nor any combination thereof, cancel in their contribution to eqs.~\eqref{eq:FindingAlphaK1} and \eqref{eq:FindingAlphaK2}.
Note, however, that one can avoid this condition if dealing with each choice of these functions individually.
Moreover, as we are only interested in the terms that provide the largest and smallest values of $q$ in these sets, in principle, it is only those terms that need to be non-vanishing.
But to simplify the analysis done here, we are including the set of all possible $q$ and so more restrictively insist that no terms shall vanish.

When this condition applies, we can include $G_2$, $G_{3\phi}$, and $G_{4\phi\phi}$ in $\mathcal{M}$.
With these final sets, we have reversed the inclusion and shown $\mathcal{M} \supset \bigcup \alpha[R^{(i)}_{\mu\nu}]$ as we have the highest elements that appear in the chains, eqs.~\eqref{eq:GammaSubsets}-\eqref{eq:GammaSubsetThree}, being included in $\mathcal{M}$.
In this case the full expression for $\mathcal{M}$ becomes 
\begin{align}
\label{eq:Malpha}
\mathcal{M} =& \alpha[\sum_{i=2}^4 R^u_{\mu\nu}] \nonumber \\
=& \alpha[G_2] \cup \alpha^{q}\alpha[G_{2\phi}] \cup \alpha^{q}\alpha[G_{4\phi}] \cup \alpha^{2q}\alpha[G_{4\phi\phi}] \cup \alpha^{2q}\alpha[G_{4X}] \nonumber\\
&\cup \alpha^{3q}\alpha[G_{5\phi\phi}]  \bigcup_{i=3,5}( \alpha^{2q}\alpha[G_{i\phi}] \cup \alpha^{3q}\alpha[G_{iX}] ).
\end{align} 
With all powers of $\alpha$ that can appear in $\mathcal{M}$ accounted for, $q_\mathcal{M}$ can easily be extracted for specified $G_i$ functions when initially defining the action.
For the term $\sum R^{(i)}_{\mu\nu}$ to vanish, as required to recover Einstein's field equations, there must be no terms independent of $\alpha$ (i.e., $\alpha^0$) in eq.~\eqref{eq:Malpha} and we must have $q_\emph{metric} < q_\mathcal{M}$ in the limit $\alpha \to \infty$, or $q_\emph{metric} > q_\mathcal{M}$ for $\alpha\to0$. These inequalities prevent divergences in the equation of motion.
Further, in order for $\Gamma \to \epsilon$, we must have $q_\emph{metric} < q_\Gamma$ ($\alpha\rightarrow\infty$) or $q_\emph{metric} > q_\Gamma$ ($\alpha\rightarrow0$) as explained above.

\subsubsection{Scalar field equation}
\label{sec:ScreeningConScalar}

In the scalar field equation \eqref{eq:HorndeskiScalarEq}, $P_\phi$, $J_\mu^{(i)}$, and $R^{(i)}_{\mu\nu}$ are functions of $\phi$.
The stress-energy tensor is also multiplied by the function $\Xi(\phi)$, which complicates the analysis since the contribution may disappear from the equation when taking a limit. This allows for the possibility of the scalar field to be sourced by self-interactions rather than $T$.
Even should the metric field equation reduce to Einstein's equations in such a limit, in order to attribute this to a screening effect, we shall require that the matter density should be the source of the scalar field equation.
This implies that the right-hand side of eq.~\eqref{eq:HorndeskiScalarEq} must not vanish (or diverge) when taking the limit $\alpha\rightarrow\infty$ (or 0), hence,
\begin{equation}
\label{eq:TScalarFactor}
\Xi \not\to 0 \,.
\end{equation}
Moreover, with a non-vanishing term on the right-hand side of eq.~\eqref{eq:HorndeskiScalarEq}, at least one term on the left must also remain to balance it. The relevant set for these conditions is 
\begin{align}
\label{eq:AlphaXi}
\alpha[\Xi] =& \alpha[G_{4\phi} + (G_{4X} - G_{5 \phi} )  \Box \phi - \frac{1}{2} G_{5X}(\Box \phi)^2] \nonumber \\
=&  \alpha[G_{4\phi}] \cup \alpha^q \alpha[G_{4X} - G_{5 \phi}] \cup \alpha^{2q} \alpha[G_{5X}]
\end{align}

In the heuristic examples given in section~\ref{sec:Method}, the first requirement was satisfied by letting a term in eq.~\eqref{eq:TScalarFactor} scale as $\alpha^0$ (such as $G_4 = \phi$) so it would not vanish when taking one of the limits.
This meant that we only had to examine the left-hand side of the equation.
For simplicity, let us in the following only consider the limit of $\alpha \to \infty$.
Analogous results, however, also apply for $\alpha \to 0$.
As both sides are now functions of $\alpha^q$, one must also consider the right-hand side and find the smallest value  of $q$ for which eq.~\eqref{eq:TScalarFactor} holds. We shall denote it as $q_\Xi$.
More specifically, for \eqref{eq:TScalarFactor}, the value of $q_\emph{scalar}$ required in the scalar field equations is $q_\emph{scalar}\leq q_\Xi$, where we have equality when eq.~\eqref{eq:TScalarFactor} contains no terms that scale as $\alpha^0$.
Next, the left-hand side of eq.~\eqref{eq:HorndeskiScalarEq} must be checked to ensure that the equation is balanced.

We first consider the second series of terms on the left-hand side of the scalar field equation~\eqref{eq:HorndeskiScalarEq}, which is 
\begin{equation}
\label{eq:RScalarFactor}
(\sum_{i=2}^5 R^{(i)}) \Xi \,. 
\end{equation}
As we have required $\sum_{i=2}^5 R^{(i)}_{\mu\nu} \to 0$  for screening in the metric field equation as well as that $\Xi$ does not diverge, eq.~\eqref{eq:RScalarFactor} vanishes when taking the limit.
We are left with the first term in \eqref{eq:HorndeskiScalarEq}, 
\begin{equation}
\label{eq:HornScalarLHS}
\left(\sum_{i=2,3} (\nabla^\mu J^{(i)}_\mu  -  P^{(i)}_\phi ) + \sum_{i=4,5} ( \overline{\nabla^\mu J_\mu^{(i)}} - \overline{P^{(i)}_\phi} )\right) \Gamma  = -\frac{T}{M_p^2} \Xi \,.
\end{equation} 
Analogous to the set $\mathcal{M}$ used in the metric field equation, let us now define the set of $\alpha$ powers
\begin{equation}
 \mathcal{S} \equiv \alpha[ \, \sum_{i=2,3} (\nabla^\mu J^{(i)}_\mu  -  P^{(i)}_\phi ) + \sum_{i=4,5} ( \overline{\nabla^\mu J_\mu^{(i)}} - \overline{P^{(i)}_\phi} )\,] \,.
\end{equation}
There are fewer unique terms in the scalar field equation than what we encountered in the metric field equation; the only one being 
\begin{equation}
G_{5X} R_{\alpha\mu \beta\nu} \nabla^\mu \nabla^\nu \phi \nabla^\alpha \nabla^\beta \phi \,.
\end{equation}
When isolating the highest terms in the chains defined in eqs.~\eqref{eq:GammaSubsets}-\eqref{eq:GammaSubsetThree}, we again insist that specific combinations contain no terms that can cancel, ensuring all possible terms remain in the final field equations. 
More specifically, for the remaining terms we impose for generality that
\begin{align}
& ( -2 G_{5 \phi \phi} + 2 G_{4\phi X} ) R_{\mu\nu} \nabla^\mu \phi \nabla^\nu \phi \,,\\
& (G_{3X} - 2 G_{4\phi X} ) (\Box \phi)^2 \,,\\
& ( 2G_{3X\phi} -2 G_{4 \phi \phi X} ) \nabla^\mu X \nabla_\mu \phi\,, \\
&(2G_{3\phi} - G_{2X})\Box \phi \,,\\
&-4G_{3\phi \phi}X - G_{2\phi}
\end{align}
contain no terms that cancel.
When satisfied, the set of all relevant powers of $\alpha$ in the scalar field equation becomes
\begin{align}
\label{eq:Salpha}
\mathcal{S} = & \alpha[ G_{2\phi} \Gamma] \cup \alpha^q \alpha[G_{2X} \Gamma] \cup \alpha^{2q} \alpha[ G_{4\phi X} \Gamma] \cup \alpha^q \alpha[G_{4X} \Gamma] \nonumber\\
 & \bigcup_{i=3,5} ( \alpha^{2q} \alpha[G_{iX} \Gamma] \cup \alpha^q \alpha[G_{i\phi} \Gamma] ) \,,
\end{align}
which allows us to determine the values $q_{\mathcal{S}}$ takes from the minimum or maximum of $\mathcal{S}$.
Then the value of $q$ in the limit of $\alpha \to \infty$ must obey $q_\emph{scalar} \leq q_{\mathcal{S}}$, where again we have equality when there are no terms independent of $\alpha$.
This inequality then ensures that the left-hand side of eq.~\eqref{eq:HorndeskiScalarEq} has at least one term that does not vanish.

\subsubsection{Einstein gravity limit}
\label{sec:EinsteinLimit}

We have formulated the requirements on the metric and scalar field equations for an Einstein gravity limit to exist, which translate directly onto conditions on $G_i$ that must be satisfied. 
This determines the powers $q_\emph{metric}$ and $q_\emph{scalar}$ that we must adopt in the scaling equation~\eqref{eq:phiExp}.
In order for the limit to be self-consistent, we further require that $q_\emph{metric} = q_\emph{scalar}$.

The metric field equation puts the only strict inequality on the value of $q_\emph{metric}$ and to recover Einstein's equations, we insisted that (see section~\ref{sec:ScreeningConMetric})
\begin{align}
 \sum R^{(i)}_{\mu \nu} \to 0 & \implies q_\emph{metric} < q_\mathcal{M} \,, \nonumber\\
 \Gamma \to \epsilon=const. & \implies q_\emph{metric} < q_\Gamma \,. \nonumber 
\end{align} 
Complimentary to those conditions, for the scalar field equation we required that (see section~\ref{sec:ScreeningConScalar})
\begin{align}
 \Xi \not \to 0  \implies q_\emph{scalar} \leq q_\Xi \,, \nonumber\\
 \left(\sum_{i=2,3} (\nabla^\mu J^{(i)}_\mu  -  P^{(i)}_\phi ) + \sum_{i=4,5} ( \overline{\nabla^\mu J_\mu^{(i)}} - \overline{P^{(i)}_\phi} )\right) \Gamma \not \to &0  \implies q_\emph{scalar} \leq q_{\mathcal{S}} \,, \nonumber
\end{align}
where the first condition guaranteed that the scalar field is sourced only by the trace of the stress-energy tensor and the second condition ensured that the right-hand side of the scalar field equation~\eqref{eq:HorndeskiScalarEq} is balanced by a contribution on the left.

In summary, for a self-consistent Einstein gravity limit, the gravitational model must satisfy the conditions:
\begin{align}
q_\emph{metric} & < q_\mathcal{M} \,,\; q_\Gamma \,, \label{eq:qCondition1}\\
q_\emph{scalar} & \leq  q_\mathcal{S} \,,\; q_\Xi \,, \label{eq:qCondition2} \\
q_\emph{metric} & = q_\emph{scalar} \,.
\end{align}
Recall that these conditions apply for the limit of $\alpha \to \infty$ and the inequalities flip when taking the limit of $\alpha\to0$ instead.
These conditions can easily be checked for a given gravitational model, and in the next section we provide a few examples.
In the case of having no terms independent of $\alpha$ in the equations of motion, the inequalities become
\begin{align}
\label{eq:qCondition3}
q_\emph{metric} & < q_\mathcal{M}\,, q_\Gamma \,, \\
\label{eq:qCondition4} 
q_\emph{scalar} & = q_\Xi =  q_\mathcal{S} \,, \\
\label{eq:qCondition5}
q_\emph{metric} & = q_\emph{scalar} \,.
\end{align}

\subsection{Examples}
\label{sec:Examples}

Let us examine a few example Lagrangians and apply the procedure laid out in sections \ref{sec:ScreeningConMetric} through \ref{sec:EinsteinLimit} to determine whether they contain a self-consistent Einstein gravity limit.
We start by re-examining the cubic galileon and chameleon models discussed in sections \ref{sec:ScalingDerivative} and \ref{sec:Cham}, respectively.

Applying eqs.~\eqref{eq:Malpha} and \eqref{eq:AlphaGamma} to the cubic galileon action~\eqref{eq:CubicAction}, we can directly identify the sets of $\alpha$ that are relevant in the the metric field equation, 
\begin{align}
\mathcal{M} &= \alpha[2\omega \phi^{-1}X] \cup \alpha^q\alpha[-2\omega \phi^{-2} X] \cup \alpha^q\alpha[1] \cup \alpha^{2q}\alpha[3\alpha\phi^{-4}X/4] \cup \alpha^{3q}\alpha[\alpha\phi^{-3}/4]\,, \\
\alpha[\Gamma] &= \alpha[\phi] = \alpha[\alpha^0\phi_0 + \alpha^q\phi_0\psi ]\,.
\end{align}
We consider the limit of $\alpha \to \infty$, for which from condition~\eqref{eq:qCondition3}, we find
\begin{equation}
q_\Gamma \in \lbrace 0 \rbrace \,, \ \ \ q_\mathcal{M} = \min \bigg\lbrace 0, -\frac{1}{4}, -\frac{1}{3} \bigg\rbrace \implies q_\emph{metric} < -\frac{1}{3} \,.
\end{equation}
From the applying eqs.~\eqref{eq:Salpha} and \eqref{eq:AlphaXi} to the action, one finds the relevant sets of $\alpha$ in the scalar field equation as
\begin{align}
\Xi =& \alpha[1] \,, \\ 
\mathcal{S} =& \alpha[-2\omega\phi^{-1}X] \cup \alpha^q\alpha[2\omega] \cup \alpha^{2q} \alpha[\alpha\phi^{-2}/4] \cup \alpha^q\alpha[-3\alpha\phi^{-3}X] \,.
\end{align}
The condition~\eqref{eq:qCondition4} implies that
\begin{equation}
q_\Xi \in \mathbb{R} \,, \ \ \ q_\mathcal{S} = \min \bigg\lbrace -\frac{1}{2} , -\frac{1}{3} \bigg\rbrace \implies q_\emph{scalar} = -\frac{1}{2} \,,
\end{equation}
where $q_\Xi$ is undetermined as there are no powers of $\alpha$ in $\Xi$.
Finally we must check for consistency between the two equations of motion, condition~\eqref{eq:qCondition5},
\begin{equation}
q_\emph{scalar} = q_\emph{metric} = - \frac{1}{2} < -\frac{1}{3} \,,
\end{equation} 
which shows that the theory contains a consistent Einstein gravity limit when $\alpha \to \infty$.
Note that we directly arrive at this conclusion from analysing the free functions $G_i$ in the action only without the need of considering the galileon equations of motion directly.

In the case of the chameleon action~\eqref{eq:ChamAction}, we let $\phi_0 = \phi_\emph{min}$ in the expansion~\eqref{eq:phiExp} which minimises the scalar field potential for $n>0$.
For the limit $\alpha\to 0$, eqs.~\eqref{eq:Malpha},~\eqref{eq:AlphaGamma} and the conditions~\eqref{eq:qCondition3} imply that
\begin{equation}
q_\Gamma \in \lbrace 0 \rbrace \,, \ \ \ q_\mathcal{M} = \max \bigg\lbrace 0, -\frac{1}{n} \bigg\rbrace \implies q_\emph{metric} > 0 \,.
\end{equation}
Examination of eqs.~\eqref{eq:Salpha},~\eqref{eq:AlphaXi} and condition~\eqref{eq:qCondition4} yields
\begin{equation}
q_\Xi \in \mathbb{R} \,, \ \ \ q_\mathcal{S} = \max \bigg\lbrace -\frac{1}{n-1} , 0 \bigg\rbrace \implies q_\emph{scalar} = -\frac{1}{n-1}
\end{equation}
when $n<1$.
For consistency between the two equations of motion, we require that 
\begin{equation}
q_\emph{scalar} = q_\emph{metric} = - \frac{1}{n-1} > 0 \,, \label{eq:chameleonconsistency}
\end{equation} 
which holds for $n\in(0,1)$ and provides an Einstein gravity limit.
In the limit of $\alpha \to \infty$, we now use the minimum of these sets and the condition~\eqref{eq:chameleonconsistency} switches signs.
Hence, we must have $ -1/(n-1) < -1/n$ and $n>1$ for a consistent Einstein gravity limit.

The most trivial example of a gravity theory without an Einstein limit is $G_4 = X$, in which case the condition \eqref{eq:G4limitCondition} is violated as $G_4$ will not tend to a constant.
Another,  more involved  example is $G_4 = \phi$, mimicking our other examples in satisfying conditions~\eqref{eq:qCondition3} and \eqref{eq:qCondition4}, but where
we further set $G_3=\alpha \phi X^{-2}$ so that $\mathcal{M}$ and $\mathcal{S}$ are non-trivial,
\begin{align}
\mathcal{M} &= \left\lbrace \alpha^{q} , \alpha^{-2q+1} , \alpha^{-3q + 1} \right\rbrace \,,\\
\mathcal{S} &= \left\lbrace \alpha^{-4q + 1} , \alpha^{ - 3 q + 1} , \alpha^{-3q + 1} , \alpha^{-2q + 1} \right\rbrace \,,
\end{align}
which follows from eqs.~\eqref{eq:Malpha} and \eqref{eq:Salpha}.
For $\alpha \to \infty$, we get $q_\mathcal{M} = 0$ and $q_\mathcal{S}=\frac{1}{4}$.
Thus, we cannot recover Einstein gravity since this would require $q_\mathcal{M} > q_\mathcal{S}$ instead.
Similarly, if we take the limit $\alpha\to0$, we obtain $q_\mathcal{M} = q_\mathcal{S} = \frac{1}{2}$, which does not satisfy the requirement for an Einstein limit, $q_\mathcal{M} < q_\mathcal{S}$, either. 

As our last example, we consider Brans-Dicke theory which possesses no screening mechanism unless we add a scalar field potential.
The model is embedded in the Horndeski action by setting $G_4 = \phi$, $G_2 = 2 \alpha \phi^{-1} X$ with all other $G_i$ vanishing. 
The usual Brans-Dicke parameter $\omega$ becomes the scaling parameter ($\omega\rightarrow\alpha$). 
We then find from eqs.~\eqref{eq:Malpha} and \eqref{eq:Salpha} that
\begin{align}
\mathcal{M} &= \lbrace \alpha^{1+2q} , \alpha^q \rbrace \,, \\
\mathcal{S} &= \lbrace \alpha^{1+q} , \alpha^{1+2q} \rbrace \,.
\end{align}
Let us first consider the case where $\alpha \to \infty$.
Then for the metric field equation to reproduce the Einstein field equations, we must have $q_{metric} < -1/2$.
The scalar field equation demands that $q_{scalar} = -1$.
Hence, conditions \eqref{eq:qCondition3} and \eqref{eq:qCondition4} are satisfied for $q=-1$ and we recover an Einstein gravity limit.
A possible recovery of GR in this model is not surprising as it is well known to succeed when $\omega$ becomes large.
However, this limit is different as in the scaling method the value of $\alpha$ is a given constant and we are taking the limits of its comparable magnitude with respect to the scalar field (see section~\ref{sec:Method}). But the example of large $\omega$ illustrates that an Einstein gravity limit may not necessarily be attributed to a screening effect. 
Considering the limit $\alpha \to 0$ instead, we find $q_\mathcal{M} = 0$ and $q_\mathcal{S} = -\frac{1}{2}$ and hence, we do not recover Einstein gravity in this limit, which would require $q_\mathcal{M}<q_\mathcal{S}$. Thus, the model provides both a limit of modified gravity and a recovery of GR whereas it is well known to not possess a screening mechanism.
We therefore need an auxiliary method to determine whether a particular Einstein gravity limit is due to a screening effect or not. 
This will be the focus of the next section.

\subsection{Radial dependence for screening}
\label{sec:RadialDependence}

When the metric field equations reduce to the Einstein field equations for a limit of $\alpha$, they become independent of the variation of the scalar field. 
The dynamics of the scalar field is then solely determined by the scalar field equation.
However, the equation of motion for the perturbation $\psi$ in the expansion~\eqref{eq:phiExp} may, in general, be a complicated differential equation.
For the expansion to be valid, we must have $\psi < \alpha^{-q}$ in the regime we wish to describe and we must check that the solution to the scalar field equation satisfies this condition. 
For the regime in question to show a recovery of Einstein gravity due to a screening effect, we furthermore want it to describe a region of high ambient density or in proximity of a massive body.

Consider, for instance, a spherical matter source.
We now must find the profile of $\psi$ around this mass to determine where the expansion breaks down.
For Vainshtein screening, we expect that the Einstein gravity limit becomes invalid once the distance from the source is large, at which point the modified gravity model becomes unscreened.
For chameleon screening, there will be a thin shell interpolating the scalar field between the minima of the effective potential set by the different ambient matter densities in the interior and exterior of the source as discussed in section~\ref{sec:Cham}, where both regions are described by the same limit of $\alpha\to0$.

Here, we outline a method for crudely approximating the radial profile $\phi(r)$ of a modified gravity model with a radial matter distribution, which can then be used to evaluate whether the Einstein gravity limits are associated with a screening mechanism or not.
As a demonstration, we apply the method to the cubic galileon with a cylindrical mass distribution and a chameleon model with a spherical mass distribution, where we show that the Einstein limit is attributed to screening.
Finally, we also consider Brans-Dicke theory with a spherical matter source and show that its Einstein limit described in section~\ref{sec:Examples} is simply associated with large distances from the mass distribution and hence is not attributed to a screening effect.
For simplicity, we shall adopt a Minkowski metric as an approximation in all scenarios.
One could also consider a Schwarzschild background, where we assume that we are describing distances large with respect to the Schwarzschild radius (but small with respect to any screening radius) so that our approximation holds. 
This assumption can easily be dropped however, and in the Einstein gravity limit, one would then be working around a non-trivial solution to the metric field equation.

In general, suppose that $\phi$ satisfies some partial differential equation
\begin{equation}
 F(\phi, \partial\phi, \partial^2\phi)= \rho M_p^{-2}.
\end{equation}
Writing out the derivatives in the coordinate choice suitable for the symmetry of the problem, such as spherical or cylindrical,  yields $F(\phi, \partial_r\phi, \partial_r^2\phi)=  \rho(r) M_p^{-2}$.
Approximating the radial derivatives as 
\begin{equation}
\partial_r \approx r^{-1} \,,
\end{equation}
and the mass distribution (for a spherical symmetry) as $\rho \approx M r^{-3}$ changes the differential equation to a polynomial equation for $\phi(r)$.
We justify this approximation on both dimensional and symmetry grounds. 
While such a mass distribution is unphysical as the integral will diverge, we are avoiding performing integrals in the approximation and so we smear out the mass source over the space we consider to get the radial dependence.
In general, there will not be an analytical solution to this equation and a numeric solution will have to be found, which is still simpler than finding a solution to the potentially nonlinear differential equation.

As our first example, consider the cubic galileon model with equation of motion
\begin{equation}
\label{eq:FlatCubic}
 \Box \phi + \alpha \left[ (\Box \phi)^2 - (\nabla_\mu \nabla_\nu \phi)^2 \right] = \frac{\rho}{M_p^2} \,.
\end{equation}
This is a simplified version of the equation of motion \eqref{eq:CubicScalarEom} that schematically contains the terms of interest.
Solutions for different symmetric mass distributions of this equation can, for instance, be found in Ref.~\citep{ShapeDependenceOfVainshteinScreening}.
Consider a cylindrical geometry and mass distribution with line element $ds^2 = -dt^2 +dr^2 + r^2 d\theta^2 + dz^2$ and scalar field profile $\phi=\phi(r)$.
The equation of motion then becomes 
\begin{equation}
\label{eq:FlatCubicRadial}
 \phi'' + \frac{\phi'}{r} +\alpha \frac{2\phi' \phi''}{r} = \frac{\rho}{M_p^2} \,,
\end{equation}
which after our approximations $\rho(r) \approx C M r^{-2}$, for some constant $C$ with units $mass$, and $\partial_r \approx r^{-1}$ becomes
\begin{equation}
 \frac{2\phi}{r^2} +\alpha \frac{2 \phi^2}{r^4} \approx \frac{ C M}{M_p^2 r^2} \,.
\end{equation}
A solution of this quadratic equation is
\begin{equation}
\phi(r) = \frac{r^2}{2 \alpha} \left(\sqrt{1+ \frac{r_v^2}{r^2}} -1 \right) \,,
\end{equation}
where we have defined the Vainshtein radius $r_v^2 \equiv 2 \alpha C M / M_p^2$.
Note that this solution differs from the exact solution found for a cylindrical top-hat mass by an overall factor of $1/2$~\citep{ShapeDependenceOfVainshteinScreening}.
However, the functional form of this simple approximation agrees with the full solution.

Applying the expansion~\eqref{eq:phiExp} with $q=-\frac{1}{2}$ and taking the limit of $\alpha \to \infty$, the equation of motion \eqref{eq:FlatCubicRadial} for the perturbation $\psi$ becomes
\begin{equation}
 \frac{2\psi' \psi''}{r} \approx \frac{2 \psi^2}{r^4} \approx \frac{C M}{M_p^2 r^2}
\end{equation}
with $\psi < \alpha^{1/2}$.
Hence, the scalar field profile is $\psi \propto r$, which needs to be small compared to $\alpha^{-q} = \alpha^{1/2} \approx r_v$.
Thus, the solution is valid within the Vainshtein radius, demonstrating that the Einstein gravity limit in the cubic galileon model can be attributed to a screening effect. 

The chameleon model of section \ref{sec:Cham} has the equation of motion
\begin{equation}
(3+2\omega)\Box \phi = V'_{eff}(\phi) = M_p^{-2} T + \alpha (\phi - \phi_m)^{n-1} (2(\phi- \phi_m) - n \phi) \,,
\end{equation}
where we will consider its approximation on Minkowski space. We showed in section \ref{sec:Examples} that when $\alpha \to 0$ and $\phi_0 \approx \phi_\emph{min}$ the equation of motion gives the solution for $\psi$ as 
\begin{equation}
\psi=\left( -\frac{M_p^{-2} T }{n\,\phi_0^n}\right)^\frac{1}{n-1}.
\end{equation}
If we approximate $T \approx -\rho \approx -M r^{-3}$, then the exponent of $r$ is positive for $n\in (0,1)$.
Hence, $\psi$ is small with respect to $\alpha^\frac{1}{1-n}$ for small $r$ and the effect can be attributed to screening in the vicinity of a source.
In contrast, for the case $n>1$ discussed in section~\ref{sec:Examples}, the exponent of $r$ is negative and the recovery of GR only occurs at large distances from the source, which does not correspond to a screening effect.

Finally, we study the example of Brans-Dicke gravity, for which in section~\ref{sec:Examples}, we found an Einstein limit when $q=-1$.
The scalar field equation can be approximated as
\begin{equation}
 \frac{1}{r^2} \psi \approx \Box \psi =  \frac{\rho}{M_p^2} \approx \frac{M}{M_p^2 r^3} \,,
\end{equation}
from which we find that $\psi \propto 1/r$.
As $\psi$ needs to be small compared to $\alpha^{-q} = \alpha$ in order for the expansion to be valid, the solution only applies to scales of $r\gtrsim \alpha M/M_p$.
Thus, the Einstein gravity limit is obtained at large distances from the matter source and cannot be attributed to a screening mechanism.
Rather, we find that far from the source, the scalar field $\phi$ decouples from the metric field equation and Einstein gravity is recovered to highest order.

A caveat with this approximate method is that it does not work for all differential equations. Take for example $\partial_r \psi = \psi$ which results in an absurdity when the approximation $\partial_r \approx \frac{1}{r}$ is made.
Furthermore, in the above approximation, we assume that the symmetry for the mass distribution is either cylindrical or spherical. 
But it is known that the morphology of the mass distribution can affect whether Vainshtein screening is operating or not (see, e.g., Ref.~\citep{ShapeDependenceOfVainshteinScreening}).
In essence, terms in the equations of motion disappear when a coordinate symmetry is imposed on the fields and hence the equations found after taking a limit of $\alpha$ may not be consistent.
The result is that the value of $q$ chosen is no longer valid as it does not provide non-vanishing terms in the equations of motion.
There may, however, still be a screening effect if, for instance, a quartic galileon term vanishes but the cubic contributes.
Conversely, our method may not indicate a screening effect with the scalar field equation becoming inconsistent despite a screening mechanism operating in the covariant equations of motion.
Further work is needed to examine these scenarios.
In this respect, it should be noted that mass distributions do not have perfect symmetry in reality. 
Even in highly symmetric cases, they will contain perturbations.
Working without covariance, one might then consider the coordinate dependence of the field to also contain a dependence on $\alpha$. 
Using the chain rule to extract powers of $\alpha$, one can then proceed with our limiting arguments.
This alternative scaling method may provide an approach to addressing morphology dependent screening, and we shall briefly outline such a method in appendix~\ref{sec:AltMethod}.

\section{Conclusion} 
\label{sec:conclusions}

In the century since Einstein's discovery of GR, a plethora of modified gravity models have been proposed to address a variety of problems in physics, ranging from the quantum nature of gravity to cosmic acceleration. 
Powerful observational tests of gravity have placed tight bounds on deviations from GR in the Solar System.
To pass these constraints and allow modifications on cosmological scales, screening mechanisms have been invoked that suppress these modifications in high-density regions or near massive bodies, where, however, small deviations from GR remain.
Screening effects are predominantly dependent on nonlinear terms in the equations of motion, making the calculation of these small deviations mathematically challenging.

To overcome some of these challenges, we introduce a method for efficiently finding the relevant equations of motion in regimes of the strong or weak coupling of extra terms in the modified gravitational action.
It works through introducing these coupling parameters to some power in an expansion of the scalar field.
The power becomes fixed when considering formal limits of the parameters, facilitating the individual study of the two opposing regimes.
This greatly aids the examination of gravity in screened regions as we can efficiently and consistently determine when terms dominate or become irrelevant in the equations of motion.
We provide two explicit examples with the cubic galileon and chameleon models, illustrating the applicability of the scaling method to both screening by derivative interactions and local field values respectively.

Thanks to the simplicity of our method, it can be applied to the general Horndeski action to find embedded models of physical interest by insisting that there exists a limit where the metric field equations become Einstein's equations. 
From this, we derive a set of conditions on the four free functions of the Horndeski action which, when satisfied, ensure this limit exists for the covariant equations of motion. 
These conditions relate both the metric and scalar field equations through the  exponent of the coupling entering in the scalar field expansion. 
Again, we illustrate the use of these conditions by re-examining the cubic galileon and chameleon models as well as two models that do not employ any screening mechanisms. 

Importantly, while an Einstein limit may exist for a given gravitational action, it is not guaranteed that it is due to a screening effect. 
To determine whether the limit can be attributed to screening, further information on the scalar field profile is required. 
By adopting an appropriate coordinate symmetry and turning the differential equation describing the dynamics of the field into a polynomial whose coefficients are functions of the radius, the scalar field profile can be approximately derived.  
If the metric field equations recover Einstein's field equations in a region of high matter density or close to a matter source, we associate the Einstein limit with a screening effect. 
We demonstrate that in the cubic galileon and chameleon models, the Einstein limits can be attributed to screening mechanisms whereas in Brans-Dicke theory, we find that the Einstein limit is associated with large distances from the matter source and so there is no screening effect. 

Importantly, our scaling method, alongside a low-energy static limit, allows for a parameterised post-Newtonian (PPN) expansion of screened theories to be performed in regimes where screening mechanisms operate. 
Previous work has preformed a PPN expansion for Vainshtein screening~\citep{TheParametrizedPostNewtonianVainshteinianFormalism} but the approach adopted is mathematically involved and has only been applied to the cubic galileon model. 
Given the simplicity of our scaling procedure, it becomes feasible to develop a PPN expansion of the Horndeski action which simultaneously takes into account the wide variety of screening mechanisms, testing all models that can be embedded in the action. 
This will be presented in separate work. 

Further, we plan to use our method to inspect gravitational waves in screened theories (see, e.g.,~\cite{GalileonRadiationFromBinarySystems,GravitationalWaveEmissionInShiftSymetricHorndeskiTheories}). 
As gravitational waves can be treated as perturbations of the metric at large distances from the source far from their sources, our method naturally compliments this analysis in theories with screening mechanisms. 
Given the recent direct measurement of gravitational waves with the LIGO detectors~\citep{ObservationofGravitationalWavesfromaBinaryBlackHoleMerger}, 
a comparison between waves in modified gravities and the observed post-Newtonian parameters~\citep{TestsOfGeneralRelativityWithGW150914} potentially allows for tests to be made on this phenomenon covering all regimes of gravity: strong and weak, screened and unscreened. 
The development of a suitable theoretical framework is of great interest and can be done in a model independent manner analogous to the PPN formalism (e.g.,~\citep{FundamentalTheoreticalBiasInGravitationalWaveAstrophysicsAndTheParametrizedPostEinsteinianFramework}) and our work may helps facilitate such an analysis. 

Finally, one may also consider the scaling method we developed outside of the context of gravity to study other classical systems that adhere to a hierarchy in the contributing terms. 

\acknowledgments

We thank Kazuya Koyama for useful discussions.
This work was supported by the STFC Consolidated Grant for Astronomy and Astrophysics at the University of Edinburgh. L.L.~also acknowledges support from a SNSF Advanced Postdoc.Mobility Fellowship (No.~161058).
Please contact the authors for access to research materials.

\newpage

\appendix

\section{Horndeski field equations}
\label{app:HornFieldEqs}

The field equations for Horndeski gravity given in Ref.~\citep{GeneralizedGInflation} are reproduced here for convenience.
Varying the action~\eqref{eq:HornAction} with respect to the metric and scalar field yields the equations of motion, 
\begin{align}
0 =& \sum_{i=2}^5 \mathcal{G}^{(i)}_{\mu\nu}, \\
0 =& \sum_{i=2}^5 (P^{(i)}_\phi - \nabla^\mu J_\mu^{(i)}),
\end{align}
respectively,
which we rearrange to get the equations~\eqref{eq:HorndeskiMetricEq} and \eqref{eq:HorndeskiScalarEq}.
Hereby, we have defined the rank-2 tensors
\begin{align}
	R^{(2)}_{\mu\nu} \equiv & -\frac{1}{2} G_{2X} \nabla_\mu \phi \nabla_\nu \phi - \frac{1}{2}g_{\mu\nu} (G_{2X} X - G_2) \,, \\
	R^{(3)}_{\mu\nu} \equiv & G_{3X} \bigg\lbrace \frac{1}{2} \Box \phi \nabla_\mu \phi \nabla_\nu \phi + \nabla_{(\mu} X \nabla_{\nu)} \phi + \frac{1}{2} g_{\mu\nu} X \Box\phi \bigg\rbrace + G_{3\phi} \nabla_\mu \phi \nabla_\nu \phi \,, \\
	R^{(4)}_{\mu\nu} \equiv & G_{4X} \bigg\lbrace -\frac{1}{2} R \nabla_\mu \phi \nabla_\nu \phi - \Box \phi \nabla_\mu \nabla_\nu \phi + \nabla_\lambda \nabla_\mu \phi \nabla^\lambda\nabla_\nu \phi + 2 R_{\lambda(\mu} \nabla_{\nu)} \phi \nabla^\lambda \phi  \nonumber\\ 
&+ R_{\mu \alpha \nu \beta} \nabla^\alpha \phi \nabla^\beta \phi- \frac{1}{2} g_{\mu\nu} ( RX + R_{\alpha \beta} \nabla^\alpha \phi \nabla^\beta \phi) \bigg\rbrace +G_{4\phi} \bigg\lbrace -\nabla_\mu \nabla_\nu \phi - \frac{1}{2} g_{\mu\nu} (\Box \phi) \bigg\rbrace \nonumber \\
&+ G_{4XX} \bigg\lbrace - \frac{1}{2} [(\Box \phi)^2 - (\nabla_\alpha \nabla_\beta \phi)^2] \nabla_\mu \phi \nabla_\nu \phi +2\nabla_\lambda X \nabla^\lambda \nabla_{(\mu} \phi \nabla_{\nu)}\phi \nonumber \\
&- \nabla_\lambda X \nabla^\lambda \phi \nabla_\mu \nabla_\nu \phi - 2 \nabla_{(\mu} X \nabla_{\nu)} \phi \Box \phi - \nabla^\alpha \phi \nabla_\alpha \nabla_\mu \phi \nabla^\beta \phi \nabla_\beta \nabla_\nu \phi  \nonumber \\
&-\frac{1}{2} g_{\mu\nu} ( X[(\Box\phi)^2 - (\nabla_\alpha \nabla_\beta \phi)^2] + 2 \nabla^\alpha X \nabla^\beta \phi \nabla_\alpha \nabla_\beta \phi - \nabla_\alpha X \nabla^\alpha \phi \Box \phi \nonumber \\
&+ (\nabla_\alpha \nabla_\lambda \phi \nabla^\alpha \phi)^2) \bigg\rbrace + G_{4\phi\phi} \bigg\lbrace - \nabla_\mu \phi \nabla_\nu \phi -\frac{1}{2} g_{\mu\nu} (-2X) \bigg\rbrace \nonumber \\
&+ G_{4X\phi} \bigg\lbrace 2 \nabla_\lambda \phi \nabla^\lambda \nabla_{(\mu} \phi \nabla_{\nu)} \phi + 2 X \nabla_\mu \nabla_\nu \phi - 2 \nabla_\mu \phi \nabla_\nu \phi \Box \phi \bigg\rbrace \,, \\
	R^{(5)}_{\mu\nu} \equiv & G_{5X} \bigg\lbrace R_{\alpha \beta} \nabla^\alpha \phi \nabla^\beta \nabla_{(\mu} \phi \nabla_{\nu)} \phi - R_{\alpha(\mu} \nabla_{\nu)} \phi \nabla^\alpha \phi \Box \phi \nonumber \\
	&- \frac{1}{2} R_{\alpha\beta} \nabla^\alpha \phi \nabla^\beta \phi \nabla_\mu \nabla_\nu \phi - \frac{1}{2} R_{\mu \alpha \nu \beta} \nabla^\alpha \phi \nabla^\beta \phi \Box \phi + R_{\alpha \lambda \beta (\mu} \nabla_{\nu)} \phi \nabla^\lambda \phi \nabla^\alpha \nabla^\beta \phi \nonumber \\
	&+ R_{\alpha \lambda \beta(\mu} \nabla_{\nu)} \nabla^\lambda \phi \nabla^\alpha \phi \nabla^\beta \phi + \nabla^\alpha X \nabla^\beta \phi R_{\alpha (\mu \nu) \beta} - \nabla_{(\mu} X G_{\nu)\lambda} \nabla^\lambda \phi \nonumber \\
	&- \nabla^\lambda X R_{\lambda (\mu} \nabla_{\nu)} \phi - \frac{1}{2} G_{\alpha \beta} \nabla^\alpha \nabla^\beta \phi \nabla_\mu \phi \nabla_\nu \phi - \frac{1}{2}\Box \phi \nabla_\alpha \nabla_\mu \phi \nabla^\alpha \nabla_\nu \phi \nonumber \\
	&+ \frac{1}{2} (\Box \phi)^2 \nabla_\mu \nabla_\nu \phi  - \frac{1}{2} g_{\mu\nu}  \bigg( R_{\alpha\beta} [ \nabla^\alpha \phi \nabla_\lambda \phi \nabla^\beta \nabla^\lambda \phi - \nabla^\alpha \phi \nabla^\beta \phi \Box \phi \nonumber \\
	&+ \nabla^\alpha X \nabla^\beta \phi ] + R_{\alpha \lambda \beta \rho} [ \nabla^\rho \nabla^\lambda \phi \nabla^\alpha \phi \nabla^\beta \phi + \nabla^\alpha X \nabla^\rho \phi g^{\lambda \beta}]  + G_{\alpha \beta} [ \nabla^\alpha \nabla^\beta \phi X \nonumber \\
	&- \nabla^\alpha X \nabla^\beta \phi ]   -\frac{1}{2} \Box \phi \nabla_\alpha \nabla_\beta \phi \nabla^\alpha \nabla^\beta \phi - \frac{1}{6} \Box \phi (\nabla_\alpha \nabla_\beta \phi)^2 \nonumber
\end{align}

\begin{align}
	&+ \frac{1}{3} (\nabla_\alpha \nabla_\beta \phi)^3 \bigg) \bigg\rbrace + G_{5\phi} \bigg\lbrace \frac{1}{2} \nabla_{(\mu} \nabla_{\nu)} \phi \Box \phi - \nabla_\lambda \nabla_{(\mu} \phi \nabla_{\nu)} \nabla^\lambda \phi + \frac{1}{2} \Box \phi \nabla_\mu \nabla_\nu \phi \nonumber \\
	&+ \nabla^\alpha \phi \nabla^\beta \phi R_{\alpha (\mu \nu) \beta} - \nabla_{(\mu} \phi G_{\nu)\lambda} \nabla^\lambda \phi - \nabla^\lambda \phi R_{\lambda (\mu} \nabla_{\nu)} \phi \nonumber \\
	&- \frac{1}{2} g_{\mu \nu} \bigg( \nabla^\alpha \phi \nabla^\beta \phi R_{\alpha \rho \lambda \beta}g^{\rho \lambda} - \frac{1}{2} R^{\alpha \beta} \nabla_\alpha \phi \nabla_\beta \phi - G_{\alpha \beta} \nabla^\alpha \phi \nabla^\beta \phi \bigg) \bigg\rbrace \nonumber \\ 
    &+ G_{5XX}  \bigg\lbrace -\frac{1}{2}\nabla_{(\mu}X \nabla^\alpha \phi \nabla_\alpha \nabla_{\nu)} \phi \Box \phi \nonumber \\
	&- \frac{1}{2}\nabla_\alpha X \nabla_\beta \phi \nabla^\alpha \nabla^\beta \phi \nabla_\mu \nabla_\nu \phi + \frac{1}{2}\nabla_{(\mu} X \nabla_{\nu)} \phi [ (\Box \phi)^2 - (\nabla_\alpha \nabla_\beta \phi)^2 ] \nonumber \\
	&+ \nabla_\alpha X \nabla_\beta \phi \nabla^\alpha \nabla_{(\mu} \phi \nabla^\beta \nabla_{\nu)}\phi - \nabla_\beta X[\Box \phi \nabla^\beta \nabla_{(\mu} \phi - \nabla^\alpha \nabla^\beta \phi \nabla_\alpha \nabla_{(\mu} \phi ] \nabla_{\nu)} \phi \nonumber \\
	&+ \frac{1}{2} \nabla^\alpha \phi \nabla_\alpha X [ \Box \phi \nabla_\mu \nabla_\nu \phi - \nabla_\beta \nabla_\mu \phi \nabla^\beta \nabla_\nu \phi] + \frac{1}{12} [ ( \Box \phi)^3 - 3 \Box \phi (\nabla_\alpha \nabla_\beta \phi)^2 \nonumber \\
	&+ 2 (\nabla_\alpha \nabla_\beta \phi)^3] \nabla_\mu \phi \nabla_\nu \phi \nonumber \\
    & - \frac{1}{2} g_{\mu \nu} \bigg( \nabla^\alpha \phi \nabla^\beta X [ \nabla^\lambda \nabla_\beta \phi \nabla_\lambda \nabla_\alpha \phi + \nabla_\beta \nabla^\lambda  \phi  \nabla_\alpha \nabla_\lambda \phi - \frac{3}{2} \nabla_\alpha \nabla_\beta \phi \Box \phi - \frac{1}{2} \nabla_\beta \nabla_\alpha \phi \Box \phi \nonumber \\
    &+ \frac{1}{2} g_{\alpha \beta} [ (\Box \phi)^2 \nonumber - (\nabla_\rho \nabla_\lambda \phi)^2 ] ]  + \nabla^\alpha X \nabla^\beta X [ \nabla_\alpha \nabla_\beta \phi - g_{\alpha \beta} \Box \phi ] - \frac{1}{6} X [ (\Box \phi)^3 \nonumber \\
	&- 3 \Box \phi (\nabla_\alpha \nabla_\beta \phi)^2 + 2 (\nabla_\alpha \nabla_\beta \phi)^3] \bigg) \bigg\rbrace + G_{5\phi X}  \bigg\lbrace  -\frac{1}{2} \nabla_{(\mu} \phi \nabla^\alpha \phi \nabla_\alpha \nabla_{\nu)} \phi \Box \phi \nonumber \\
	&+ \frac{1}{2}\nabla_{(\mu} X \nabla_{\nu)} \phi \Box \phi - \nabla_\lambda X \nabla_{(\mu} \nabla_{\nu)} \nabla^\lambda \phi + \frac{1}{2}[ \nabla_\lambda X \nabla^\lambda \phi - \nabla_\alpha \phi \nabla_\beta \phi \nabla^\alpha \nabla^\beta \phi ] \nabla_\mu \nabla_\nu \phi \nonumber \\
	&+ \frac{1}{2} \nabla_\mu \phi \nabla_\nu \phi [ (\Box \phi)^2 - (\nabla_\alpha \nabla_\beta \phi)^2 ] + \nabla_\alpha \phi \nabla_\beta \phi \nabla^\alpha \nabla_{(\mu} \phi \nabla^\beta \nabla_{\nu)} \phi - \nabla_\beta \phi [ \Box \phi \nabla^\beta \nabla_{(\mu} \phi \nonumber \\
	&- \nabla^\alpha \nabla^\beta \phi \nabla_\alpha \nabla_{(\mu} \phi ] \nabla_{\nu)} \phi - X [ \Box \phi \nabla_\mu \nabla_\nu \phi - \nabla_\beta \nabla_\mu \phi \nabla^\beta \nabla_\nu \phi] \nonumber \\
	&- \frac{1}{2} g_{\mu \nu} \bigg( \nabla^\alpha \phi \nabla^\beta \phi [ -2 \nabla_\alpha \nabla_\beta \phi \Box \phi + \frac{1}{2}g_{\alpha\beta} [ (\Box \phi)^2 - (\nabla_\rho \nabla_\lambda \phi)^2] ] \nonumber \\
	&+ \nabla^\alpha \phi \nabla^\beta X [ - 2 g_{\alpha \beta} \Box \phi + \nabla_\alpha \nabla_\beta \phi] \bigg) \bigg\rbrace  \nonumber \\
	&+G_{5\phi\phi}\bigg\lbrace \frac{1}{2} \nabla_\mu \phi \nabla_\nu \phi \Box \phi - \nabla_\lambda \phi \nabla_{(\mu} \phi \nabla_{\nu)} \nabla^\lambda \phi - X \nabla_\mu \nabla_\nu \phi \bigg\rbrace \,.
\end{align}
To simplify the scalar field equation, we have defined the scalars
\begin{align}	
P^{(2)}_\phi \equiv & G_{2\phi} \,, \\
P^{(3)}_\phi \equiv & \nabla_\mu G_{3\phi} \nabla^\mu \phi \,, \\
P^{(4)}_\phi \equiv & G_{4\phi} R + G_{4 \phi X} [ (\Box \phi)^2 - (\nabla_\mu \nabla_\nu \phi)^2], \\
P^{(5)}_\phi \equiv & -\nabla_\mu G_{5 \phi} G^{\mu\nu}\nabla_\nu \phi - \frac{1}{6}G_{5\phi X} [(\Box \phi)^3 - 3 \Box \phi (\nabla_\mu \nabla_\nu \phi)^2 + 2(\nabla_\mu \nabla_\nu \phi)^3 ]
\end{align}
and the covariant four-vectors
\begin{align}
J^{(2)}_\mu \equiv  & - \mathcal{L}_{2X} \nabla_\mu \phi \,, \\
J^{(3)}_\mu \equiv  & - \mathcal{L}_{3X} \nabla_\mu \phi + G_{3X} \nabla_\mu X + 2 G_{3\phi} \nabla_\mu \phi \,, \\
J^{(4)}_\mu  \equiv & - \mathcal{L}_{4X} \nabla_\mu \phi + 2 G_{4X} R_{\mu\nu} \nabla^\nu \phi -   2 G_{4XX}(\Box \phi \nabla_\mu X - \nabla^\nu X \nabla_\mu \nabla_\nu \phi) \nonumber \\
	&- 2 G_{4\phi X} (\Box \phi \nabla_\mu \phi + \nabla_\mu X) \,, \\
J^{(5)}_\mu \equiv  & - \mathcal{L}_{5X} \nabla_\mu \phi - 2 G_{5\phi} G_{\mu \nu} \nabla^\nu \phi - G_{5X} [ G_{\mu \nu}\nabla^\nu X + R_{\mu\nu} \Box \phi \nabla^\nu \phi - R_{\nu\lambda} \nabla^\nu \phi \nabla^\lambda \nabla_\mu \phi \nonumber \\
	&- R_{\alpha \mu \beta \nu} \nabla^\nu \phi \nabla^\alpha \nabla^\beta \phi ] + G_{5XX} \bigg\lbrace \frac{1}{2} \nabla_\mu X [(\Box \phi)^2 - (\nabla_\alpha \nabla_\beta \phi)^2]  \nonumber \\
	&- \nabla_\nu X (\Box \phi \nabla_\mu \nabla^\nu \phi - \nabla_\alpha \nabla_\mu \phi \nabla^\alpha \nabla^\nu \phi) \bigg\rbrace + G_{5\phi X} \bigg\lbrace \frac{1}{2} \nabla_\mu \phi [ (\Box \phi)^2 - (\nabla_\alpha \nabla_\beta \phi)^2] \nonumber \\
	&+ \Box \phi \nabla_\mu X - \nabla^\nu X \nabla_\nu \nabla_\mu \phi \bigg\rbrace \,,
\end{align}
where we have used the different components of the Horndeski Lagrangian
\begin{align}
\mathcal{L}_2 \equiv  &  G_2(\phi,X) \,, \\
\mathcal{L}_3 \equiv  & -G_3(\phi,X)\Box\phi \,, \\
\mathcal{L}_4 \equiv  & G_4(\phi,X)R+ G_{4X}(\phi,X)[(\Box \phi)^2 - (\nabla_\mu \nabla_\nu \phi)^2] \,, \\
\mathcal{L}_5 \equiv  & G_5(\phi, X)G_{\mu\nu} \nabla^\mu \nabla^\nu \phi - \frac{G_{5X}(\phi,X)}{6}[(\Box \phi)^3 - 3 \Box \phi (\nabla_\mu \nabla_\nu \phi)^2 + 2 (\nabla_\mu \nabla_\nu \phi)^3] \,.
\end{align}

\section{Coordinate-dependent scaling method}
\label{sec:AltMethod}

We briefly present here another method that uses the limiting argument developed in section~\ref{sec:MyMethod}, but where $\alpha$ enters as a scaling of the coordinates instead.
This method is most useful when applied to theories with derivative interactions such as galileon models. 

For simplicity, let the scalar field be a function of one variable only $\phi = \phi(r)$ that scales as $\phi(\alpha^q r)$ with $q$ a real number.
Note, however, that one can easily generalise the following results to also include a dependence, for instance, on angular coordinates.
We again consider a generic scalar field equation of the form 
\begin{equation}
\label{eq:GenericFCoord}
\alpha^s F_u(\phi, \partial_r\phi, \partial_r^2\phi) + \alpha^t F_v(\phi, \partial_r\phi, \partial_r^2\phi) = \rho \,,
\end{equation}
for some functions $F_u, F_v$ and real numbers $u, v$. After a redefinition of the coordinates, eq.~\eqref{eq:GenericFCoord} becomes
\begin{equation}
\alpha^s F_u(\phi, \alpha^q \partial_{\alpha^q r}\phi, \alpha^{2q} \partial_{\alpha^q r}^2\phi) + \alpha^t F_v(\phi, \alpha^q \partial_{\alpha^q r}\phi, \alpha^{2q} \partial_{\alpha^q r}^2\phi) = \rho \,.
\end{equation}
We now put the condition on both $F_u$ and $F_v$ that $\alpha^q$ is factorized out with order their subscript, which yields the exponents $s+uq$ and $t+vq$ of $\alpha$.
The values of $q$ must ensure a term independent of $\alpha$ on the left-hand side, so that
\begin{equation}
q \in \left\lbrace -\frac{s}{u}, -\frac{t}{v} \right\rbrace = Q \,.
\end{equation}
In order to prevent divergences, we must adopt the most negative value in $Q$ when $\alpha \to \infty$ and the most positive when $\alpha \to 0$, respectively.
The resulting equations for $\phi$, should they exist, describe the field in both limits.

For a specific example, consider a cubic galileon model with the scalar field equation~\eqref{eq:FlatCubic} in approximately flat space.
Assuming a cylindrical mass distribution $\rho = \rho (r)$ and field profile $\phi = \phi(r)$, the equation of motion becomes
\begin{equation}
\label{eq:ClyindricalEOM}
\frac{\rho(r)}{M_p} = \partial_r^2 \phi+ \frac{\partial_r \phi}{r} + \frac{2 \alpha \partial_r \phi \partial_r^2 \phi}{r} \,.
\end{equation}
To find the leading term in the limit of $\alpha\to\infty$, we insist that the coordinate dependence of the field goes as $\alpha^q r$ such that upon applying the chain rule we get
\begin{equation}
\frac{\rho(r)}{M_p} = \alpha^{2q} \partial_{\alpha^q r}^2 \phi+ \alpha^{q}\frac{\partial_{\alpha^q r} \phi}{r} + 2 \alpha^{1+3q}\frac{ \partial_{\alpha^q r} \phi \partial_{\alpha^q r}^2 \phi}{r} \,,
\end{equation}
from which we conclude that $q=-\frac{1}{3}$.
Hence, we arrive at the simple scalar field equation
\begin{equation}
\frac{r\rho(r)}{M_p} = 2 \partial_{\alpha^{-1/3} r} \phi \partial_{\alpha^{-1/3} r}^2 \phi = \partial_{\alpha^{-1/3} r}(\partial_{\alpha^{-1/3} r}\phi)^2 \,.
\end{equation}
Inside of a cylindrical mass where $\rho(r) = \rho_0$, then this is trivially integrated to
\begin{equation}
\phi = \sqrt{\frac{\rho_0}{4M_p}}r^2 \alpha^{3q/2} =  \sqrt{\frac{\rho_0}{4M_p}}r^2 \alpha^{-1/2} \,,
\end{equation}
where we have insisted that $\phi \to 0$ as $r\to 0$.
Notice that we again recover that there is an overall factor of $\alpha^{-1/2}$, which agrees with what we found with the method described in section \ref{sec:ScalingDerivative}.
An advantage of this alternative scaling method, in contrast to the scaling method introduced in section \ref{sec:MyMethod}, is that one can adopt a different scaling for each coordinate direction, and so encode the morphological dependence of the screening mechanism into the limiting procedure.
A disadvantage, however, is that the method is computationally more involved, requiring the equations of motion to be written for a given coordinate choice. 
With that we also lose the benefits of the covariant method, making an analysis like the one performed in section~\ref{sec:GRHorndeski} infeasible.

\bibliographystyle{JHEP} 

\bibliography{Scaling_The_Horndeski_Action}

\providecommand{\href}[2]{#2}\begingroup\raggedright\begin{thebibliography}{10}

\bibitem{ObservationalEvidenceFromSupernovaeForAnAcceleratingUniverseAndACosmologicalConstant}
A.~G. {Riess}, A.~V. {Filippenko}, P.~{Challis}, A.~{Clocchiatti},
  A.~{Diercks}, P.~M. {Garnavich}, R.~L. {Gilliland}, C.~J. {Hogan}, S.~{Jha},
  R.~P. {Kirshner}, B.~{Leibundgut}, M.~M. {Phillips}, D.~{Reiss}, B.~P.
  {Schmidt}, R.~A. {Schommer}, R.~C. {Smith}, J.~{Spyromilio}, C.~{Stubbs},
  N.~B. {Suntzeff}, and J.~{Tonry}, {\it {Observational Evidence from
  Supernovae for an Accelerating Universe and a Cosmological Constant}},  {\em
  The Astronomical Journal} {\bf 116} (Sept., 1998) 1009--1038,
  [\href{http://arxiv.org/abs/astro-ph/9805201}{{\tt astro-ph/9805201}}].

\bibitem{MeasurementsOfOmegaAndLambdaFrom42HighRedshiftSupernovae}
S.~{Perlmutter}, G.~{Aldering}, G.~{Goldhaber}, R.~A. {Knop}, P.~{Nugent},
  P.~G. {Castro}, S.~{Deustua}, S.~{Fabbro}, A.~{Goobar}, D.~E. {Groom}, I.~M.
  {Hook}, A.~G. {Kim}, M.~Y. {Kim}, J.~C. {Lee}, N.~J. {Nunes}, R.~{Pain},
  C.~R. {Pennypacker}, R.~{Quimby}, C.~{Lidman}, R.~S. {Ellis}, M.~{Irwin},
  R.~G. {McMahon}, P.~{Ruiz-Lapuente}, N.~{Walton}, B.~{Schaefer}, B.~J.
  {Boyle}, A.~V. {Filippenko}, T.~{Matheson}, A.~S. {Fruchter}, N.~{Panagia},
  H.~J.~M. {Newberg}, W.~J. {Couch}, and T.~S.~C. {Project}, {\it {Measurements
  of {$\Omega$} and {$\Lambda$} from 42 High-Redshift Supernovae}},  {\em The
  Astrophysical Journal} {\bf 517} (June, 1999) 565--586,
  [\href{http://arxiv.org/abs/astro-ph/9812133}{{\tt astro-ph/9812133}}].

\bibitem{Planck2015ResultsXIIICosmologicalParameters}
{Planck Collaboration}, P.~A.~R. {Ade}, N.~{Aghanim}, M.~{Arnaud},
  M.~{Ashdown}, J.~{Aumont}, C.~{Baccigalupi}, A.~J. {Banday}, R.~B.
  {Barreiro}, J.~G. {Bartlett}, and et~al., {\it {Planck 2015 results. XIII.
  Cosmological parameters}},  {\em ArXiv e-prints} (Feb., 2015)
  [\href{http://arxiv.org/abs/1502.01589}{{\tt arXiv:1502.01589}}].

\bibitem{TheQuantumVacuumAndTheCosmologicalConstantProblem}
S.~E. {Rugh} and H.~{Zinkernagel}, {\it {The Quantum Vacuum and the
  Cosmological Constant Problem}},  {\em ArXiv High Energy Physics - Theory
  e-prints} (Dec., 2000) [\href{http://arxiv.org/abs/hep-th/0012253}{{\tt
  hep-th/0012253}}].

\bibitem{TheCosmologicalConstantProblem}
S.~Weinberg, {\it The cosmological constant problem},  {\em Rev. Mod. Phys.}
  {\bf 61} (Jan, 1989) 1--23.

\bibitem{ObservationOfANewBosonAtAMassOf125GeVWithTheCMSexperimentAtTheLHC}
S.~{Chatrchyan}, V.~{Khachatryan}, A.~M. {Sirunyan}, A.~{Tumasyan}, W.~{Adam},
  E.~{Aguilo}, T.~{Bergauer}, M.~{Dragicevic}, J.~{Er{\"o}}, C.~{Fabjan}, and
  et~al., {\it {Observation of a new boson at a mass of 125 GeV with the CMS
  experiment at the LHC}},  {\em Physics Letters B} {\bf 716} (Sept., 2012)
  30--61, [\href{http://arxiv.org/abs/1207.7235}{{\tt arXiv:1207.7235}}].

\bibitem{TheContentOffRGravity}
A.~{Nunez} and S.~{Solganik}, {\it {The content of f(R) gravity}},  {\em ArXiv
  High Energy Physics - Theory e-prints} (Mar., 2004)
  [\href{http://arxiv.org/abs/hep-th/0403159}{{\tt hep-th/0403159}}].

\bibitem{SecondOrderScalarTensorFieldEquationsInAFourDimensionalSpace}
G.~W. {Horndeski}, {\it {Second-Order Scalar-Tensor Field Equations in a
  Four-Dimensional Space}},  {\em International Journal of Theoretical Physics}
  {\bf 10} (Sept., 1974) 363--384.

\bibitem{FromKEssenceToGeneralizedGalileons}
C.~{Deffayet}, X.~{Gao}, D.~A. {Steer}, and G.~{Zahariade}, {\it {From
  k-essence to generalized Galileons}},  {\em Phys. Rev. D} {\bf 84} (Sept.,
  2011) 064039, [\href{http://arxiv.org/abs/1103.3260}{{\tt arXiv:1103.3260}}].

\bibitem{GeneralizedGInflation}
T.~{Kobayashi}, M.~{Yamaguchi}, and J.~{Yokoyama}, {\it {Generalized
  G-Inflation --- Inflation with the Most General Second-Order Field Equations
  ---}},  {\em Progress of Theoretical Physics} {\bf 126} (Sept., 2011)
  511--529, [\href{http://arxiv.org/abs/1105.5723}{{\tt arXiv:1105.5723}}].

\bibitem{4DGravityOnABraneIn5DMinkowskiSpace}
G.~{Dvali}, G.~{Gabadadze}, and M.~{Porrati}, {\it {4D gravity on a brane in 5D
  Minkowski space}},  {\em Physics Letters B} {\bf 485} (July, 2000) 208--214,
  [\href{http://arxiv.org/abs/hep-th/0005016}{{\tt hep-th/0005016}}].

\bibitem{ResummationofMassiveGravity}
C.~{de Rham}, G.~{Gabadadze}, and A.~J. {Tolley}, {\it {Resummation of Massive
  Gravity}},  {\em Physical Review Letters} {\bf 106} (June, 2011) 231101,
  [\href{http://arxiv.org/abs/1011.1232}{{\tt arXiv:1011.1232}}].

\bibitem{BeyondTheCosmologicalStandardModel}
A.~{Joyce}, B.~{Jain}, J.~{Khoury}, and M.~{Trodden}, {\it {Beyond the
  cosmological standard model}},  {\em Physics Reports} {\bf 568} (Mar., 2015)
  1--98, [\href{http://arxiv.org/abs/1407.0059}{{\tt arXiv:1407.0059}}].

\bibitem{GalileonAsALocalModificationOfGravity}
A.~{Nicolis}, R.~{Rattazzi}, and E.~{Trincherini}, {\it {Galileon as a local
  modification of gravity}},  {\em Phys. Rev. D} {\bf 79} (Mar., 2009) 064036,
  [\href{http://arxiv.org/abs/0811.2197}{{\tt arXiv:0811.2197}}].

\bibitem{HealthyTheoriesBeyondHorndeski}
J.~{Gleyzes}, D.~{Langlois}, F.~{Piazza}, and F.~{Vernizzi}, {\it {New Class of
  Consistent Scalar-Tensor Theories}},  {\em Physical Review Letters} {\bf 114}
  (May, 2015) 211101, [\href{http://arxiv.org/abs/1404.6495}{{\tt
  arXiv:1404.6495}}].

\bibitem{TransformingGravityTheoriesBeyondTheHorndeskiLagrangian}
M.~{Zumalac{\'a}rregui} and J.~{Garc{\'{\i}}a-Bellido}, {\it {Transforming
  gravity: From derivative couplings to matter to second-order scalar-tensor
  theories beyond the Horndeski Lagrangian}},  {\em Physics Review D} {\bf 89}
  (Mar., 2014) 064046, [\href{http://arxiv.org/abs/1308.4685}{{\tt
  arXiv:1308.4685}}].

\bibitem{DegenerateHigherDerivativeTheoriesBeyondHorndeski}
D.~{Langlois} and K.~{Noui}, {\it {Degenerate higher derivative theories beyond
  Horndeski: evading the Ostrogradski instability}},  {\em Journal of Cosmology
  and Astroparticle Physics} {\bf 2} (Feb., 2016) 034,
  [\href{http://arxiv.org/abs/1510.06930}{{\tt arXiv:1510.06930}}].

\bibitem{TheConfrontationBetweenGeneralRelativityAndExperiment}
C.~M. {Will}, {\it {The Confrontation between General Relativity and
  Experiment}},  {\em Living Reviews in Relativity} {\bf 17} (June, 2014) 4,
  [\href{http://arxiv.org/abs/1403.7377}{{\tt arXiv:1403.7377}}].

\bibitem{ModifiedGravityAndCosmology}
T.~{Clifton}, P.~G. {Ferreira}, A.~{Padilla}, and C.~{Skordis}, {\it {Modified
  gravity and cosmology}},  {\em Phys. Rep.} {\bf 513} (Mar., 2012) 1--189,
  [\href{http://arxiv.org/abs/1106.2476}{{\tt arXiv:1106.2476}}].

\bibitem{Cosmologicaltestsofmodifiedgravity}
K.~{Koyama}, {\it {Cosmological tests of modified gravity}},  {\em Reports on
  Progress in Physics} {\bf 79} (Apr., 2016) 046902,
  [\href{http://arxiv.org/abs/1504.04623}{{\tt arXiv:1504.04623}}].

\bibitem{DarkEnergyVsModifiedGravity}
A.~{Joyce}, L.~{Lombriser}, and F.~{Schmidt}, {\it {Dark Energy vs. Modified
  Gravity}},  {\em ArXiv e-prints} (Jan., 2016)
  [\href{http://arxiv.org/abs/1601.06133}{{\tt arXiv:1601.06133}}].

\bibitem{BreakingADarkDegeneracyWithGravitationalWaves}
L.~{Lombriser} and A.~{Taylor}, {\it {Breaking a dark degeneracy with
  gravitational waves}},  {\em Journal of Cosmology and Astroparticle Physics}
  {\bf 3} (Mar., 2016) 031, [\href{http://arxiv.org/abs/1509.08458}{{\tt
  arXiv:1509.08458}}].

\bibitem{ChallengestoSelfAccelerationinModifiedGravity}
L.~{Lombriser} and N.~A. {Lima}, {\it {Challenges to Self-Acceleration in
  Modified Gravity}},  {\em ArXiv e-prints} (Feb., 2016)
  [\href{http://arxiv.org/abs/1602.07670}{{\tt arXiv:1602.07670}}].

\bibitem{ChameleonFieldsAwaitingSurprisesForTestsOfGravityInSpace}
J.~{Khoury} and A.~{Weltman}, {\it {Chameleon Fields: Awaiting Surprises for
  Tests of Gravity in Space}},  {\em Physical Review Letters} {\bf 93} (Oct.,
  2004) 171104, [\href{http://arxiv.org/abs/astro-ph/0309300}{{\tt
  astro-ph/0309300}}].

\bibitem{ScreeningLongRangeForcesthroughLocalSymmetryRestoration}
K.~{Hinterbichler} and J.~{Khoury}, {\it {Screening Long-Range Forces through
  Local Symmetry Restoration}},  {\em Physical Review Letters} {\bf 104} (June,
  2010) 231301, [\href{http://arxiv.org/abs/1001.4525}{{\tt arXiv:1001.4525}}].

\bibitem{ToTheProblemOfNonvanishingGravitationMass}
A.~I. {Vainshtein}, {\it {To the problem of nonvanishing gravitation mass}},
  {\em Physics Letters B} {\bf 39} (May, 1972) 393--394.

\bibitem{KMOUFLAGEGravity}
E.~{Babichev}, C.~{Deffayet}, and R.~{Ziour}, {\it {k-MOUFLAGE Gravity}},  {\em
  International Journal of Modern Physics D} {\bf 18} (Dec., 2009) 2147--2154,
  [\href{http://arxiv.org/abs/0905.2943}{{\tt arXiv:0905.2943}}].

\bibitem{ClassifyingLinearlyShieldedModifiedGravityModelsInEffectiveFieldTheory}
L.~{Lombriser} and A.~{Taylor}, {\it {Classifying Linearly Shielded Modified
  Gravity Models in Effective Field Theory}},  {\em Physical Review Letters}
  {\bf 114} (Jan., 2015) 031101, [\href{http://arxiv.org/abs/1405.2896}{{\tt
  arXiv:1405.2896}}].

\bibitem{ClassicalDualsOfDerivativelySelfCoupledTheories}
G.~{Gabadadze}, K.~{Hinterbichler}, and D.~{Pirtskhalava}, {\it {Classical
  duals of derivatively self-coupled theories}},  {\em Phys. Rev. D} {\bf 85}
  (June, 2012) 125007, [\href{http://arxiv.org/abs/1202.6364}{{\tt
  arXiv:1202.6364}}].

\bibitem{ClassicalDualsLegendreTransformsAndTheVainshteinMechanism}
A.~{Padilla} and P.~M. {Saffin}, {\it {Classical duals, Legendre transforms and
  the Vainshtein mechanism}},  {\em Journal of High Energy Physics} {\bf 7}
  (July, 2012) 122, [\href{http://arxiv.org/abs/1204.1352}{{\tt
  arXiv:1204.1352}}].

\bibitem{CosmologyOnABraneInMinkowskiBulk}
C.~{Deffayet}, {\it {Cosmology on a brane in Minkowski bulk}},  {\em Physics
  Letters B} {\bf 502} (Mar., 2001) 199--208,
  [\href{http://arxiv.org/abs/hep-th/0010186}{{\tt hep-th/0010186}}].

\bibitem{Ghostsintheselfacceleratingbraneuniverse}
K.~{Koyama}, {\it {Ghosts in the self-accelerating brane universe}},  {\em
  Physical Review D} {\bf 72} (Dec., 2005) 123511,
  [\href{http://arxiv.org/abs/hep-th/0503191}{{\tt hep-th/0503191}}].

\bibitem{MoreOnGhostsInTheDvaliGabadazePorratiModel}
D.~{Gorbunov}, K.~{Koyama}, and S.~{Sibiryakov}, {\it {More on ghosts in the
  Dvali-Gabadaze-Porrati model}},  {\em Physics Review D} {\bf 73} (Feb., 2006)
  044016, [\href{http://arxiv.org/abs/hep-th/0512097}{{\tt hep-th/0512097}}].

\bibitem{ClassicalAndQuantumConsistencyOfTheDGPModel}
A.~{Nicolis} and R.~{Rattazzi}, {\it {Classical and Quantum Consistency of the
  DGP Model}},  {\em Journal of High Energy Physics} {\bf 6} (June, 2004) 59,
  [\href{http://arxiv.org/abs/hep-th/0404159}{{\tt hep-th/0404159}}].

\bibitem{StrongInteractionsAndStabilityInTheDGPModel}
M.~A. {Luty}, M.~{Porrati}, and R.~{Rattazzi}, {\it {Strong interactions and
  stability in the DGP model}},  {\em Journal of High Energy Physics} {\bf 9}
  (Sept., 2003) 029, [\href{http://arxiv.org/abs/hep-th/0303116}{{\tt
  hep-th/0303116}}].

\bibitem{CovariantGalileon}
C.~{Deffayet}, G.~{Esposito-Far{\`e}se}, and A.~{Vikman}, {\it {Covariant
  Galileon}},  {\em Phys. Rev. D} {\bf 79} (Apr., 2009) 084003,
  [\href{http://arxiv.org/abs/0901.1314}{{\tt arXiv:0901.1314}}].

\bibitem{WeaklyBrokenGalileonSymmetry}
D.~{Pirtskhalava}, L.~{Santoni}, E.~{Trincherini}, and F.~{Vernizzi}, {\it
  {Weakly broken galileon symmetry}},  {\em Journal of Cosmology and
  Astroparticle Physics} {\bf 9} (Sept., 2015) 007,
  [\href{http://arxiv.org/abs/1505.00007}{{\tt arXiv:1505.00007}}].

\bibitem{ObservationofGravitationalWavesfromaBinaryBlackHoleMerger}
B.~P. {Abbott}, R.~{Abbott}, T.~D. {Abbott}, M.~R. {Abernathy}, F.~{Acernese},
  K.~{Ackley}, C.~{Adams}, T.~{Adams}, P.~{Addesso}, R.~X. {Adhikari}, and
  et~al., {\it {Observation of Gravitational Waves from a Binary Black Hole
  Merger}},  {\em Physical Review Letters} {\bf 116} (Feb., 2016) 061102,
  [\href{http://arxiv.org/abs/1602.03837}{{\tt arXiv:1602.03837}}].

\bibitem{TheParametrizedPostNewtonianVainshteinianFormalism}
A.~{Avilez-Lopez}, A.~{Padilla}, P.~M. {Saffin}, and C.~{Skordis}, {\it {The
  Parametrized Post-Newtonian-Vainshteinian formalism}},  {\em Journal of
  Cosmology and Astroparticle Physics} {\bf 6} (June, 2015) 044,
  [\href{http://arxiv.org/abs/1501.01985}{{\tt arXiv:1501.01985}}].

\bibitem{TheoryAndExperimentInGravitationalPhysics}
C.~M. {Will}, {\em {Theory and Experiment in Gravitational Physics}}.
\newblock Mar., 1993.

\bibitem{MachsPrincipleAndARelativisticTheoryOfGravitation}
C.~{Brans} and R.~H. {Dicke}, {\it {Mach's Principle and a Relativistic Theory
  of Gravitation}},  {\em Physical Review} {\bf 124} (Nov., 1961) 925--935.

\bibitem{HaloModellingInChameleonTheories}
L.~{Lombriser}, K.~{Koyama}, and B.~{Li}, {\it {Halo modelling in chameleon
  theories}},  {\em Journal of Cosmology and Astroparticle Physics} {\bf 3}
  (Mar., 2014) 021, [\href{http://arxiv.org/abs/1312.1292}{{\tt
  arXiv:1312.1292}}].

\bibitem{ShapeDependenceOfVainshteinScreening}
J.~K. {Bloomfield}, C.~{Burrage}, and A.-C. {Davis}, {\it {Shape dependence of
  Vainshtein screening}},  {\em Phys. Rev. D} {\bf 91} (Apr., 2015) 083510,
  [\href{http://arxiv.org/abs/1408.4759}{{\tt arXiv:1408.4759}}].

\bibitem{GalileonRadiationFromBinarySystems}
C.~{de Rham}, A.~{Matas}, and A.~J. {Tolley}, {\it {Galileon radiation from
  binary systems}},  {\em Phys. Rev. D} {\bf 87} (Mar., 2013) 064024,
  [\href{http://arxiv.org/abs/1212.5212}{{\tt arXiv:1212.5212}}].

\bibitem{GravitationalWaveEmissionInShiftSymetricHorndeskiTheories}
E.~{Barausse} and K.~{Yagi}, {\it {Gravitation-Wave Emission in Shift-Symmetric
  Horndeski Theories}},  {\em Physical Review Letters} {\bf 115} (Nov., 2015)
  211105, [\href{http://arxiv.org/abs/1509.04539}{{\tt arXiv:1509.04539}}].

\bibitem{TestsOfGeneralRelativityWithGW150914}
{The LIGO Scientific Collaboration} and {the Virgo Collaboration}, {\it {Tests
  of general relativity with GW150914}},  {\em ArXiv e-prints} (Feb., 2016)
  [\href{http://arxiv.org/abs/1602.03841}{{\tt arXiv:1602.03841}}].

\bibitem{FundamentalTheoreticalBiasInGravitationalWaveAstrophysicsAndTheParametrizedPostEinsteinianFramework}
N.~{Yunes} and F.~{Pretorius}, {\it {Fundamental theoretical bias in
  gravitational wave astrophysics and the parametrized post-Einsteinian
  framework}},  {\em Physical Review D} {\bf 80} (Dec., 2009) 122003,
  [\href{http://arxiv.org/abs/0909.3328}{{\tt arXiv:0909.3328}}].

\end{thebibliography}\endgroup

\end{document}